\documentclass[11pt]{article}
\pdfoutput=1
\usepackage[utf8]{inputenc}
\usepackage[T1]{fontenc}

\usepackage{formatting}
\usepackage{shortcuts}

\begin{document} 
\begin{titlepage}
\begin{center}
\phantom{ }
\vspace{1cm}

{\bf \Large{Spread complexity and the saturation of wormhole size}}
\vskip 0.5cm
Vijay Balasubramanian${}^{\dagger,1,2,3}$, Javier M. Magan${}^{\ddagger, 4}$, Poulami Nandi${}^{*,1}$, Qingyue Wu${}^{**,1}$
\vskip 0.05in

\small{${}^{1}$ \textit{David Rittenhouse Laboratory, University of Pennsylvania}}
\vskip -.4cm
\small{\textit{ 209 S.33rd Street, Philadelphia, PA 19104, USA}}

\vskip -.10cm
\small{${}^{2}$ \textit{Rudolf Peierls Centre for Theoretical Physics, University of Oxford}}
\vskip -.4cm
\small{\textit{Beecroft Building, Parks Road Oxford OX1 3PU, UK}}

\vskip -.10cm
\small{  ${}^{3}$ \textit{Theoretische Natuurkunde, Vrije Universiteit Brussel}}
\vskip -.4cm
\small{\textit{Pleinlaan 2,  B-1050, Brussels, Belgium}}

\vskip -.10cm
\small{  ${}^{4}$ \textit{Instituto Balseiro, Centro At\'omico Bariloche}}
\vskip -.4cm
\small{\textit{ 8400-S.C. de Bariloche, R\'io Negro, Argentina}}

\begin{abstract}
Recent proposals equate the size of Einstein-Rosen bridges in JT gravity to spread complexity of a dual, double-scaled SYK theory (DSSYK).  We show that the auxiliary ``chord basis'' of these proposals is an extrapolation  from a sub-exponential part of the finite-dimensional physical Krylov basis of a spreading thermofield double state. The physical tridiagonal Hamiltonian coincides with the DSSYK approximation on the initial Krylov basis, but deviates markedly over an exponentially large part of the state space. We non-perturbatively extend the identification of ER bridge size and  spread complexity to the complete Hilbert space, and show that it saturates at late times. We use methods for tridiagonalizing random Hamiltonians to study all universality classes to which large N SYK theories and JT gravities can belong. The  saturation dynamics depends on the universality class, and displays ``white hole'' physics at late times where the ER bridge shrinks from maximum size to a plateau. We describe extensions of our results to higher dimensions.

\end{abstract}
\end{center}

\small{\vspace{1 cm}\noindent ${}^{\dagger}$vijay@physics.upenn.edu\\
${}^{\ddagger}$javier.magan@cab.cnea.gov.ar \\
${}^{*}$ pnandi@sas.upenn.edu\\
${}^{**}$ aqwalnut@sas.upenn.edu
}

\end{titlepage}

\setcounter{tocdepth}{2}

{\parskip = .4\baselineskip \tableofcontents}
\newpage

\section{Introduction}\label{I}

A general rule of the AdS/CFT correspondence is that any gauge-invariant quantity on one side of the duality should also be computable from the theory on the other side.  For example, consider an Einstein-Rosen (ER) bridge, i.e., an eternal black hole, dual to the thermofield double state in two CFTs \cite{Maldacena:2001kr}.  The volume of the ER bridge measured on a maximal spatial slice anchored at specific times on the two spacetime boundaries is independent of the coordinates we use, and hence is gauge-invariant.  In the classical solution this volume keeps increasing forever, beyond the timescale of  thermalization in the dual field theory at which standard observables equilibrate.  This tension, along with evidence from a series of thought experiments, led the authors of \cite{Susskind:2014rva,Stanford:2014jda} to conjecture that the volumes of Einstein-Rosen (ER) bridges codify the ``quantum complexity'' of  global time evolution of the dual thermofield double state.\footnote{See \cite{Brown:2015bva,Belin:2021bga} for alternate proposals concerning  geometric duals of quantum complexity.} 
To make this conjecture precise we must define what we mean by ``complexity''.  For example, we might mean the circuit complexity of the time evolving state \cite{Brown:2016wib,Jefferson:2017sdb, Chapman:2017rqy, Magan:2018nmu,Caputa:2018kdj, Balasubramanian:2018hsu}, or we might adopt Nielsen's geometric definition \cite{Nielsen:2005mkt, Dowling:2006tnk, Nielsen:2006cea} in terms of geodesic lengths in an appropriately defined complexity metric on the unitary group \cite{Balasubramanian:2019wgd,Bueno:2019ajd, Balasubramanian:2021mxo, Brown:2022phc}. These definitions all reproduce the long-term linear growth of the ER bridge volume. But, in all these definitions,  complexity growth saturates at times exponentially large in the entropy, essentially because a system with a finite number of degrees of freedom has an upper bound on how complicated it can be. If quantum complexity in these CFTs is related to ER bridge volumes, it follows that the bridges must stop growing in size at exponentially large times.  This saturation of ER bridges is not visible in classical General Relativity.

Recently, the authors of \cite{Balasubramanian:2022gmo,Balasubramanian:2022lnw} proposed 
a possible quantum mechanism for ER bridge saturation.  They constructed a basis for the microstates of eternal black holes in AdS consisting of ER bridges of different lengths supported internally by shells of matter. These ``bridge states'' are perturbatively orthogonal, and naively give rise to an infinite space of possible black hole microstates with ever-growing lengths.  However, small overlaps arise from the quantum effects of topology changing wormholes in the gravitational path integral.  These  overlaps make the shell states linearly dependent and force them to span  a Hilbert space with the expected Bekenstein-Hawking entropy. In particular, we can pick a complete basis of ER bridges of bounded length.  This means that  arbitarily long ER bridges, of the kind that appear at late time slices of the classical geometry, can be expanded as a superposition of short bridges, implying a sort of quantum mechanical saturation.  This mechanism parallels the saturation of the thermal two-point functions in JT gravity \cite{Saad:2019pqd} due to quantum tunnelling to baby universes with shorter ER bridges. Of course, if a large  ER bridge should be thought of as a superposition of small  bridges, we should also ask whether there is any definite notion of volume which can be identified with complexity in the dual theory.
In any case, these arguments suggest that part of the problem put forward in \cite{Susskind:2014rva,Stanford:2014jda} is related to the construction of finite dimensional Hilbert spaces with the correct Bekenstein-Hawking dimensionality  for black holes in quantum gravity. This problem has seen significant development in recent years \cite{Penington:2019kki,Hsin:2020mfa,Chandra:2022fwi,Balasubramanian:2022gmo,Balasubramanian:2022lnw,Boruch:2023trc,Antonini:2023hdh,Climent:2024trz,Iliesiu:2024cnh,Boruch:2024kvv,Balasubramanian:2024yxk}. In particular, the black microstates described in \cite{Balasubramanian:2022gmo,Balasubramanian:2022lnw,Climent:2024trz,Iliesiu:2024cnh} are helpful for our purpose since they are directly labeled by the size of the ER bridges they contain.

Dynamically speaking, imagine that we start out well-localized at $t=0$ on one of the short ER bridge microstates of  \cite{Balasubramanian:2022gmo,Balasubramanian:2022lnw}. Then the fact that the classical dynamics makes the bridge grow longer tells us that the quantum state should spread to acquire support on microstates with longer bridges.  This widening of the wavefunction can be quantified by the spread complexity \cite{SpreadC} of states,  a notion inspired by the operator growth analysis of \cite{Parker2018AHypothesis} (see \cite{Nandy:2024htc} for a  review).  Spread complexity is the minimal measure of the spread of a wavefunction over the Hilbert space as time evolves, and as such constitutes a sensible physical notion of complexity \cite{SpreadC}. The authors of \cite{Balasubramanian:2022gmo} used this scenario to set up a simple model with a Hamiltonian hopping between bridge microstates in which the spread complexity equals the expected value of the ER bridge volume, and both saturate when the wavefunction is delocalized across the complete, finite basis of microstates. A more precise example arises in the Double-Scaled-Sachdev-Ye-Kitaev model (DSSYK) \cite{Erd_s_2014,Cotler:2016fpe,Berkooz_2018,Berkooz:2018jqr}.\footnote{Another precise example of the connection between spread complexity and gravity has appeared recently in \cite{Caputa:2024sux,Fan:2024iop}.}  The authors of \cite{Berkooz_2018} solved the DSSYK model by a method that involved computing moments of the Hamiltonian.  To do so they introduced an auxiliary ``chord basis Hilbert space'', in which the Hamiltonian was tridiagonalized, and calculated all quantities in the double scaling limit in which this chord Hilbert space is inifinite dimensional. They further showed that this auxiliary Hilbert space has a bulk description in terms of Jackiw-Teitelboim (JT) gravity \cite{Berkooz_2018}.  Interestingly, it was later understood that in this double-scaled description the ER bridge volume in JT is precisely the spread complexity \cite{Lin:2022rbf,Rabinovici:2023yex}.

However, this formulation raises some important problems. First, in the strict double scaling limit in which these computations were carried out, the auxiliary Hilbert space is infinite dimensional. So we do not see late time saturation of either the spread complexity or the bulk volume that we might expect in the physical setting away from this limit. Second, the physical status of the auxiliary chord Hilbert space is unclear from the discussion. Is it an auxiliary Hilbert space or is it related to the real Hilbert space of the underlying SYK theory? Since the connection between spread complexity and the ER bridge appears in the chord Hilbert space, this is a further important question to answer.

To solve both problems,  we will  construct the tridiagonalized Hamiltonian for large but finite dimensional SYK models. In our construction the chord basis is explicitly physical -- more precisely, it coincides with an orthonormal basis for a sub-exponential part of the microscopic Hilbert space that is explored in the semiclassical limit, and at sufficiently early times in the time evolution of the thermofield double state of the full quantum theory.  Our construction exploits recent results on the  tridiagonalization of Random Matrix Theories \cite{Balasubramanian:2022dnj}, and is a special case of the methods developed in \cite{SpreadC}.  At sub-exponential times, the results of \cite{Berkooz_2018,Lin:2022rbf,Rabinovici:2023yex} show that the spread complexity of the evolving TFD state in the large N SYK theory equals the size of the dual semiclassical ER bridge.  We extrapolate this identification between spread complexity and ER bridge size to the full quantum Hilbert space, and find that the bridge size saturates at late times.
The main input to our calculations is the density of states derived by Erdos and Schr\"{o}der \cite{Erd_s_2014}.

Apart from the density of states, we find that the precise dynamics also depends on the  quantum chaotic universality class. In fact, it is known \cite{Stanford:2019vob} that the SYK model for different numbers of fermions, and Jackiw-Teilboim (JT) gravity with different non-perturbative completions, produce theories with spectra matching the three  Dyson and  seven Altland-Zirnbauer  random matrix ensembles. These ensembles define  different chaotic universaliy classes.  Using methods from \cite{Balasubramanian:2022dnj,Balasubramanian:2023kwd}, we demonstrate how these universality classes affect the late time properties of the ER bridges in the gravity theory dual to  DSSYK. We will show that the saturation value of the the ER bridge does not depend on the universality class, as expected from the analysis in \cite{Balasubramanian:2023kwd}. But the approach to saturation does strongly depend on the Dyson $\beta$ index, and mildly on the parameter $\alpha$ characterizing the Altland-Zirnbauer ensembles. In parallel to results in \cite{SpreadC,Balasubramanian:2022dnj,Erdmenger:2023shk},  we find the ER bridge grows to a larger size than the saturating value and then bounces back, a gravitational example of the phenomenon called the ``complexity slope'' in \cite{SpreadC}. As we will discuss, in this quantum gravity context the complexity slope can be interpreted as a transition to ``white hole''  dynamics, recalling recent work on the emergence of firewalls at late times in black holes \cite{Stanford:2022fdt}. 

We finish by describing how our techniques and results go beyond the DSSYK scenario. In particular we describe how they can be applied to thermofield double theories with general temperatures and to more realistic models of quantum gravity, such as holographic theories in higher dimensions. These techniques then provide a novel method for reconstructing finite dimensional Hilbert spaces that account for black hole entropy, complementing recent approaches in \cite{Penington:2019kki,Hsin:2020mfa,Chandra:2022fwi,Balasubramanian:2022gmo,Balasubramanian:2022lnw,Boruch:2023trc,Antonini:2023hdh,Climent:2024trz,Iliesiu:2024cnh}.  Our approach also relates the problem of explaining the Bekenstein-Hawking entropy to  the version of the black hole information paradox concerning late time behavior of the spectral form factor \cite{Maldacena:2001kr,Cotler:2016fpe}.

\paragraph{Plan of the article.} In Sec.~\ref{II} we review  background material: (a) the  SYK model, the tridiagonal Hamiltonian approach to solving the model in the double scaled limit \cite{Berkooz_2018}, and the triple-scaled limit where one recovers JT gravity;  (b) an introduction to the  RMT ensembles pertinent to SYK models and JT gravity; and (c) the notion of spread complexity. In Sec.~\ref{III} we construct finite dimensional tridiagonal matrices which reproduce the moments of the Hamiltonian of the DSSYK model for initial thermofield double (TFD) states. Using these tridiagonal matrices we analyze the approach to saturation of the spread complexity for different quantum chaotic universality classes that arise from variants of the SYK model and JT gravity.  At early times the spread complexity is precisely equal to the volume of the dual semiclassical ER bridge, and we extrapolate to late times to take spread complexity as the quantum definition of ER bridge volume.  We find the late time stationary value of this volume is almost equal to the exponential of the thermodynamic entropy, but somewhat larger. Sec.~\ref{IV} describes how to extend these techniques to higher dimensions and discusses the appearance of white hole physics at late times. It also discusses the relation between the Lanczos approach for thermofield double evolution and the construction of the Hilbert space of black hole microstates in quantum gravity. We end in Sec.~\ref{V} with a summary and discussion of open questions.

\paragraph{Node added.} While this paper was being completed, we received \cite{Nandy:2024zcd} which has partial overlap with our construction of a finite dimensional tridiagonal Hamiltonian accounting for the DSYK density of states.

\section{Preliminary background}\label{II}

This section contains no new results. It sets up the concepts and conventions used in the rest of the article. The reader can jump ahead to the next section and come back if needed.

\subsection{SYK, DSSYK and JT}
\label{sec:sykReview}
The Sachdev-Ye-Kitaev model \cite{sachdev,kitaev} is simultaneously a solvable many-body  interacting quantum system and a model of 2d quantum black holes \cite{kitaev,sachdev,sachdev2,Polchinski:2016xgd,Maldacena:2016hyu,Jensen:2016pah,maldacena2016conformal,PhysRevD.94.126010,Kitaev:2017awl,Sarosi:2017ykf}.  The model contains $N$ Majorana fermions $\psi_i$, $i=1,\cdots,N$,\footnote{We normalize them as
$
\left\lbrace \psi_i,\psi_j  \right\rbrace = \psi_i\psi_j+\psi_j\psi_i=2\delta_{ij}
$.}
interacting via the Hamiltonian
\be{}
\label{hamilsyk}
H=i^{p/2}\sum\limits_{1\leq i_1<i_2<\cdots <i_p\leq N}\,J_{i_1i_2\cdots i_p}\,\psi_{i_{1}}\psi_{i_{2}}\cdots \psi_{i_{p}}\;.
\ee
The couplings $J$ are real, independently distributed, Gaussian random variables with the  moments
\be 
\langle J_{i_1 i_2 \hdots i_p} \rangle =0\,\,\,\,\,\,\,\,\,\,\,\,\,
\langle J^2_{i_1 i_2 \hdots i_p} \rangle=(1-e^{-\lambda}){\begin{pmatrix}
N\\
p
\end{pmatrix} }^{-1} \mathcal{J}^2\;.
\label{eq:SYKcouplings}
\ee
where $\lambda=\frac{2p^2}{N}$. The scaling of the couplings is tuned so that the density of states is bounded between $-2\mathcal{J}$ and $2\mathcal{J}$ in the large $N$ limit.
The dimension of the Hilbert space is $L=2^{N/2}$. Below we will normalize physical quantities by this dimension. This model has been studied extensively in the literature in the large-$N$ limit with $p$ fixed (see \cite{Sarosi:2017ykf} for a review).

A cousin of this model, the quantum spin glass, was analyzed by the authors of \cite{Erd_s_2014} in a double scaling limit with $N,p \to \infty$ and $\lambda=\frac{2p^2}{N}$  fixed, in terms of ``chord diagrams'', which we will adapt and use below.  
In the same limit, the authors of \cite{Cotler:2016fpe} determined the exact low energy density of states of the SYK model.\footnote{This density of states was also found by different methods in \cite{Stanford:2017thb,Mertens_2017,Kitaev:2017awl}.} This triple-scaled limit of the SYK model (large $N$, large $p$, low $E$) is described by a Schwarzian effective action \cite{Berkooz_2018,Berkooz:2018jqr} and can by analyzed by the chord diagram method of \cite{Erd_s_2014}.

\begin{figure}
    \includegraphics[width=0.3\linewidth]{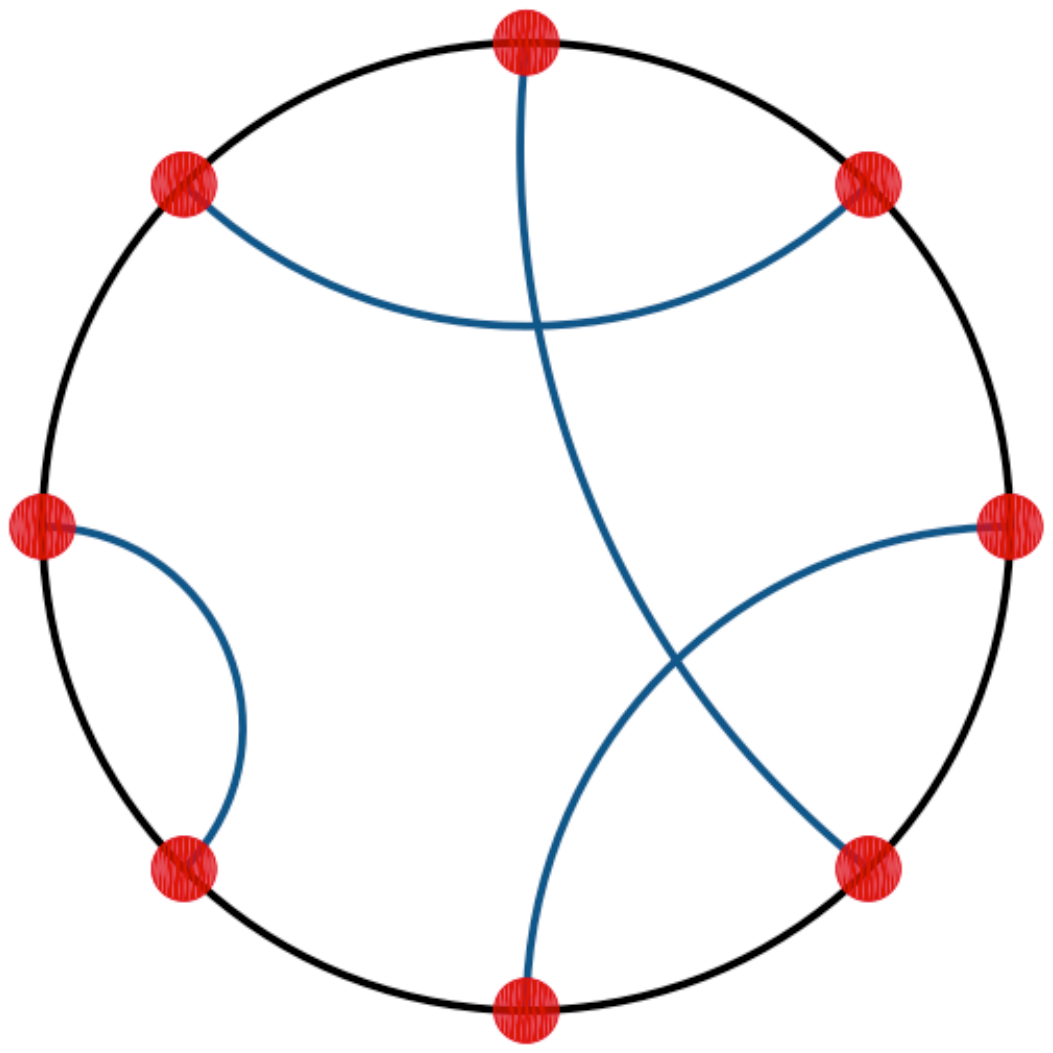}
    \centering
    \caption{An example of a chord diagram. The nodes correspond to Hamiltonian insertions. The edges connecting nodes pairwise correspond to Gaussian contractions between the SYK couplings. The circle indicates this diagram is computing a trace. In the DSSYK limit, these diagrams are the only ones contributing to the Hamiltonian moments (\ref{momtr}).}\label{chords2}
\end{figure}

We will review the chord diagram method used in \cite{Erd_s_2014} to find the exact density of states, applied directly to the SYK model instead of the original quantum spin glass.  To find the density of states, we will start by evaluating its moments.  In other words, we will compute the ensemble averaged moments of the SYK Hamiltonian \eqref{hamilsyk}:
\be{}
\label{momtr}
M_{2k}\equiv \langle \textrm{Tr} (H^{2k}) \rangle = i^{kp}\sum_{I_1,\cdots, I_{2k} } \langle J_{I_1}\cdots J_{I_{2k}} \rangle\cdot \textrm{Tr} (\psi_{I_1}\cdots \psi_{I_{2k}})\;,
\ee
where we have defined $\psi_I=\psi_{i_1}\psi_{i_2}\cdots \psi_{i_p}$. The trace is normalized so that $\textrm{Tr}(1)=1$. The odd moments vanish by gaussianity of the couplings.   We will depict the computation diagrammatically by a circle with nodes marked on it corresponding to insertions of $H$; thus $H^{2k}$ is represented by $2k$ nodes. There are products of Hamiltonians with different couplings $J_{i_1\: i_2\: \hdots i_p}$ inside the trace \eqref{momtr}, and we will draw chords between nodes on the circle to describe pairwise Wick contractions among these couplings (Fig.~\ref{chords2}).  This is a ``chord diagram''.

A chord between nodes indicate that two sets of $p$ unordered distinct sites $i_1,\cdots i_p$ (denoted by the monomials $\psi_I$) are identical. The next step is to untangle the chords so that contracted nodes are adjacent  -- in other words we want to commute operators within the trace until contracted pairs are adjcent. The authors of \cite{Erd_s_2014} showed that the number of commutations to untangle a chord diagram equals the number of intersection of the chords, and each commutation gives a factor of $q\equiv e^{-\lambda}$ in the doubled scaled limit.  Thus, the moments obtained from the chord diagrams can be  expressed as
\be\label{chordm}
M_{2k}=(1-q)^k\mathcal{J}^{2k} \sum_{\text{chord diagrams 
 with k chords}} q^{\text{count of interactions}}.
\ee
Next \cite{Erd_s_2014} uses previous results \cite{Riordan1975TheDO,10.1007/978-3-662-04166-6_17,ISMAIL1987379} to demonstrate that (\ref{chordm}) are the  moments of the distribution
\be \label{rd}
\rho(E\vert q)=\frac{2}{\pi \sqrt{4\mathcal{J}^2-E^2}}\prod\limits_{k=0}^{\infty}\left[\frac{1-q^{2k+2}}{1-q^{2k+1}}\left(1-\frac{q^k E^2}{\mathcal{J}^2(1+q^k)^2}\right)  \right] \;.
\ee
So (\ref{rd}) is the desired density of states. Furthermore, we can write the moments explicitly as 
\be \label{moman}
M_{2k}=\mathcal{J}^{2k}\sum\limits_{j=-k}^{k}(-1)^j\,q^{\frac{ j(j-1)}{2}}\binom{2k}{k+j}\;.
\ee

The authors of \cite{Berkooz_2018} described a different way to arrive at the density of states and associated moments by introducing an auxiliary infinite  dimensional Hilbert space $\mathcal{H}$ motivated by the chord diagrams.  By cutting open the diagram they viewed the computation of a Hamiltonian moment as a transition amplitude. We associate at an arbitrary point along the circle of the chord diagram with an initial ``no chord'' state of an auxiliary Hilbert space.  As we move along the circle we cross nodes.  If the node is attached to a chord that we have not crossed before we say that we have transition to a state with one more chord.  If pass a node connected to a chord that had previously  crossed, we say that we transition to a state with one less chord.  Thus we define an auxiliary Hilbert space spanned by an infinite but countable set of orthonormal states
\be 
\vert l\rangle\,\,\,\,\,\,\,\,\,\,l=,0,1,2,\cdots\,\,\;.
\ee
One of the main outcomes of \cite{Berkooz_2018} is then to show that the original moments of the SYK Hamiltonian in the double scaling limit can be encoded in the auxiliary Hilbert space as a transition amplitude of the form
\be \label{trimo}
M_{2k}= \langle \textrm{Tr} (H^{2k}) \rangle = \langle 0\vert\,T^{2k}\,\vert 0\rangle\;,
\ee
with a tridiagonal  transfer matrix $T$ of the form
\be\label{transf}
T=\mathcal{J}\begin{pmatrix}
0 & \sqrt{1-q} & 0 & 0 & \hdots\\
\sqrt{1-q}  & 0 & \sqrt{1-q^2} & 0  & \hdots\\
0 &  \sqrt{1-q^2} & 0 & \sqrt{1-q^3}  & \hdots\\
0 & 0 & \sqrt{1-q^3} &0 & \hdots\\
\vdots & \vdots & \vdots & \vdots & \ddots
\end{pmatrix}\; .
\ee
See \cite{Berkooz:2018jqr,Lin:2022rbf,Rabinovici:2023yex} for further developments and applications. An advantage of this approaches is that one can compute correlation functions exactly \cite{Berkooz:2018jqr}, by using an underlying quantum group structure to diagonalize the transfer matrix.

This approach raises some obvious questions. Does the chord Hilbert space have any physical meaning? The authors of \cite{Berkooz_2018} and \cite{Rabinovici:2023yex} showed that at low energies it is natural to adopt a continuum limit  in which the chord Hilbert space $l$ becomes continuous, and in this limit the transfer matrix $T$ becomes the Liouville Hamiltonian of JT gravity \cite{Harlow:2018tqv}.
This result was used in \cite{Rabinovici:2023yex} to show that the length of the ER bridge in these limits equals the spread complexity of \cite{SpreadC}, a quantity that we will review below. This suggests that the chord Hilbert space is not an arbitrary auxiliary construction, but is actually a presentation of the physical Hilbert space of the theory.  On the other hand, this identification is in obvious tension with the fact that the fundamental Hilbert space of an SYK model is discrete and finite dimensional. Of course the analyses of \cite{Berkooz_2018,Rabinovici:2023yex} were being performed in the thermodynamic limit in which the Hilbert space is infinite dimensional, but as in Random Matrix Theory, there should be finite size effects informing us that the Hilbert space is discrete and finite away from this limit.   Therefore, while mathematically convenient, the physical status of the chord Hilbert space remains obscure at a fundamental level. One of the objectives of this article is to settle this issue.

\subsection{Random Matrix Ensembles}
\label{sec:SYK-RMT}

Our strategy for making progress is to construct a tractable Hamiltonian that reproduces the necessary features of the SYK theory.  First, we need a Hamiltonian with the SYK density of states (\ref{rd}).  Second, we would like to reproduce the known correlations in the spectral statistics of SYK-like models.  It was shown in \cite{You:2016ldz} that when the number of fermions $N>>1$ in an SYK model is $0$ mod 8, the spectrum contains the correlations of matrices drawn from the Gaussian Orthogonal Ensemble (GOE).  Likewise if $N \mod 8 = 2,6$, the SYK model is in the Gaussian Unitary Ensemble (GUE) universality class.  Finally if $N \mod 8 = 4$ the theory belongs to the Gaussian Symplectic Ensemble (GSE) universality class.  Thus all three Dyson RMT ensembles \cite{osti_4801180} appear as effective descriptions of classes of SYK models at large $N$. From another perspective, \cite{stanford2020jt} showed that the partition sums of variants of JT gravity match the partition sums of the three Dyson RMT ensembles \cite{osti_4801180} and the seven Altland-Zirnbauer ensembles \cite{alt_zirn_orig}. Thus, our strategy will be to use the methods of \cite{SpreadC,Balasubramanian:2022dnj,Balasubramanian:2023kwd} to construct large, but finite dimensional Hamiltonians that reproduce the density of states in (\ref{rd}), along with the spectral statistics of the  Dyson and Altland-Zirbauer RMT ensembles.

To this end, we now review the relevant features of Random Matrix Theory (RMT); see \cite{Meh2004,Guhr:1997ve,bookhaake,akemann2011oxford} for extensive accounts. The main objective is to set the conventions we will use through the rest of the article. An RMT is defined as an ensemble of $L\times L$ matrices $H_{ij}$ with a distribution controlled by a potential $V(H)$
\be 
p(H)=\frac{1}{Z_{\beta_D}}e^{-\frac{\beta_D L}{4} \,\textrm{Tr}(V(H))}\;.
\ee
The matrices in the ensemble are self-adjoint. Their entries can be either real, complex, or quaternions. These correspond respectively to Dyson indices $\beta_D=1$, $2$, or $4$. The joint probability distribution for the eigenvalues of these ensembles is
\be \label{jointb}
p(\lambda_1,\cdots ,\lambda_n)=Z_{\beta_D,L}\, e^{-\frac{\beta_D\,L}{4}\sum_{k}V(\lambda_k)}\,\prod_{i<j}\vert \lambda_i-\lambda_j\vert^{\beta_D}\;.
\ee
This follows from a change of variables to the diagonal form.\footnote{In this form a natural generalization is to take $\beta_D$ to any positive value. Below we will only analyze the standard  $\beta_D=1$, $2$, or $4$. For an analysis of spread complexity with real values of $\beta_D$ see \cite{Balasubramanian:2023kwd}.} The potential and the associated density of states are related through the saddle point relation
\be \label{sadrh}
\frac{1}{4}V'(\omega)=\text{p.v.}\int \frac{\rho(E)}{\omega-E}dE,
\ee
where p.v. indicates that we take the principal value of the integral. Notice that knowing the potential or the density of states does not define the ensemble completely. One further needs to choose a universality class, or equivalently, the Dyson index.  Apart from the Dyson cases parametrized by  $\beta_D$, we also have the seven Altland-Zirnbauer universality classes \cite{osti_4801180}. These are defined by the following joint probability distributions for the eigenvalues
\be
\rho(\lambda)=Z_{\alpha, \beta, L}\, e^{-\frac{\beta L}{4}\sum_{k}V(\lambda_k)}\prod_{i<j} |\lambda_i^2-\lambda_j^2|^\beta \prod_i |\lambda_i|^\alpha \;,\label{alpha_beta}
\ee
where $L$ is the number of distinct eigenvalues in the spectrum (both positive and negative). 
For potentials of the form $V(x)\propto x^2$, these ensembles can be simulated by Gaussian random matrices with block structure described in Table~\ref{AZ_table}, which also includes the requisite values of $\alpha, \beta$.

\begin{table}[]
\begin{tabular}{l|l|l|ll}
Symmetry & $\alpha$ & $\beta$ & block structure &  \\ \hline
$U_1$   & N/A & 2 & Complex $M^\dagger=M$ &\\ 
$O_1$   & N/A & 1 & Real $M^T=M$ &  \\
$S_1$   & N/A & 4 & Quaternion $M^\dagger=M$ & \\
$\begin{bmatrix}U_1&0\\0 &U_2\end{bmatrix}$  & 1 & 2& $\begin{bmatrix}0&Z\\Z^\dagger &0\end{bmatrix}$, Complex $Z$ &\\
$S_1=\exp \begin{bmatrix}X&Y\\-\overline{Y} &\overline{X}\end{bmatrix}$& 2 & 2& $\begin{bmatrix}A&B\\\bar{B} &-\bar{A}\end{bmatrix}$, Complex $A^\dagger = A$, Complex $B^T=B$ &\\
$O_1$&  0 & 2 & Complex $M^\dagger=M$, $M^T=-M$ (i.e. imaginary)&\\
$\begin{bmatrix}O_1&0\\0 &O_2\end{bmatrix}$ & 0 & 1 & $\begin{bmatrix}0&A\\A^T &0\end{bmatrix}$, Real $A$ &\\
$\begin{bmatrix}U_1&0\\0 &\overline{U}_1\end{bmatrix}$& 1 & 1 & $\begin{bmatrix}0&\bar{Z}\\Z &0\end{bmatrix}$, Complex $Z^T=Z$ &\\
$\begin{bmatrix}U_1&0\\0 &\overline{U}_1\end{bmatrix}$       &   1 & 4& $\begin{bmatrix}0&\bar{Y}\\-Y &0\end{bmatrix}$, Complex $Y^T=-Y$ &\\
$\begin{bmatrix}S_1&0\\0 &S_2\end{bmatrix}$           &           3 & 4& $\begin{bmatrix}0&B\\B^\dagger &0\end{bmatrix}$, Quaternion $B$ &
\end{tabular}
\caption{The Dyson and Altland-Zirnbauer random matrix ensembles, along with their $\alpha, \beta$ indices and block properties. The first column lists the structures of the unitaries whose conjugation operation does not affect the distribution of random matrices. $O_i, U_i, S_i$ are matrices in the orthogonal, unitary, and quaternion unitary groups, respectively. A derivation is provided in Appendix~\ref{appendixA}. The second and third columns correspond to the exponents $\alpha,\beta$ in \eqref{alpha_beta}. The last column consists of block constructions of elements of the ensembles, collected from tables in Appendix~A and B of \cite{good_table_ref}.  These are also found in \cite{stanford2020jt,tenfold_way,alt_zirn_orig}\label{AZ_table}.
}
\end{table}

\subsection{Spread complexity}

We are going to show that ``chord bases'' of the kind used to solve the DSSYK model in \cite{Erd_s_2014,Berkooz_2018} coincide with the sub-exponential, early parts of Krylov bases for thermofield double states (TFD) evolving according to model Hamiltonians that have the correct density of states and spectral statistics to describe the various SYK and JT universality classes.  We will also discuss the size of the ER bridge appearing in the gravity description of these systems in relation to the spread complexity \cite{SpreadC} of the TFD state in the model Hamiltonian \cite{Lin:2022rbf,Rabinovici:2023yex}.

To this end, we now review the notion of spread complexity put forward in \cite{SpreadC}. In quantum mechanics, time evolution of a state is generated by the Hamiltonian as
\be\label{evoq}
\vert \psi (t)\rangle= e^{-iHt}\vert \psi (0)\rangle\;.
\ee
We want to characterize the ``complexity'' of this evolution. A natural expectation is that more complex time evolution will spread  $\vert \psi(t) \rangle$ more widely over the Hilbert space. But the extent of the spread depends on the choice of basis. In the spirit of other notions of quantum complexity, \cite{SpreadC} proposed to minimize the spread over all choices of basis. The resulting notion is was dubbed ``spread complexity''.

Surprisingly, it turns out that this minimization over Hilbert space bases can be done for all times when the  evolution is in discrete  steps, and for (at least) a finite amount of time for continuous evolution. In the continuous case, which is the one of interest below, a canonical basis follows from a formal power series expansion of (\ref{evoq}):
\be\label{exp}
\vert \psi (t)\rangle=\sum^\infty_{n=0}\frac{(-it)^n}{n!}\vert \psi_n\rangle ~~~~;~~~~ \vert \psi_n\rangle=H^n\vert \psi(0) \rangle  \, .
\ee 
Applying the Gram–Schmidt procedure to $\vert\psi_n\rangle$  we obtain an ordered, orthonormal basis
$\mathcal{K}=\set{\ket{K_n}: n=0,1,2,\cdots}$. The basis $\mathcal{K}$ is called the Krylov basis in the recent literature and turns out to minimize the spread of the wavefunction over a finite amount of time \cite{SpreadC}. It also maximizes the classicality of the state \cite{Basu:2024tgg}. When applied to operator evolution, the notion of spread complexity recovers the Krylov complexity proposal put forward in \cite{Parker2018AHypothesis}. See \cite{Nandy:2024htc} for a  review and references.

Following the details of the Gram–Schmidt procedure one concludes that in the Krylov basis the Hamiltonian is tridiagonal
\be\label{triH}
T\equiv H_{\textrm{Krylov}}= \begin{pmatrix}
a_0 & b_1 &  & & & \\
b_1 & a_1 & b_2 & & &\\
& b_2 & a_2 & b_3 & & \\
& & \ddots & \ddots & \ddots & \\
& & & b_{N-2} & a_{N-2} & b_{N-1}\\
&  & &  &b_{N-1} & a_{N-1}
\end{pmatrix}\;.
\ee
In this basis the Hamiltonian can be pictured as describing a one-dimensional chain with nearest-neighbor hopping parameters given by  $b_n$, and with the initial state on the first site on the chain. The sequences $a_n,b_n$ terminate at the dimension $N$ of the explored Hilbert space. The sequences $a_n,b_n$ are called the Lanczos coefficients or the Lanczos spectrum. This tridiagonal form is also known as a ``Hessenberg form'' of a Hermitian matrix. It is not unique, and depends on the choice of the first vector in the new basis. This Lanczos spectrum is a functional (and vice versa) of the moments of the Hamiltonian in the initial state
\be
\mu_n =
\langle K_0 |\, (iH)^n \,| K_0 \rangle \, ,
\label{eq:momgenfn}
\ee
a result that follows from  examination of the survival amplitude $\langle K_0 |\, e^{-itH} \,| K_0 \rangle$, namely the amplitude for the state to remain unchanged \cite{SpreadC}. This way of calculating the Lanczos spectrum is useful because the moment method remains valid for infinite-dimensional systems, even when direct tridiagonalization is not possible.  Note also that the computation of the density of states in the DSSYK model starts from similar Hamiltonian moments (\ref{momtr}), a connection that we will exploit below. Analytical computations of Lanczos spectra are sparse in the literature; see \cite{viswanath2008recursion,doi:10.1063/1.533010,PhysRevD.47.1640} for older examples and \cite{Parker2018AHypothesis,Berkooz_2018,Caputa:2021sib,SpreadC,Muck:2022xfc,Balasubramanian:2022dnj,Caputa:2023vyr} for more recent ones.

Intuitively, the Krylov basis is sort of in the middle between the ``computational basis'' in which the Hamiltonian was usually constructed and the energy eigenbasis that diagonalizes the Hamiltonian.  By tridiagonalizing the Hamiltonian we have made progress toward diagonalization but did not go the whole way. In fact, this was  the original motivation for the work of Lanczos \cite{Lanczos1950AnIM} --  the Krylov basis retains some of the physicality of the computational basis while giving simpler description of time evolution. The energy eigenbasis, on the other hand, fully solves the time evolution but we typically lose physical intuition for the basis.

Finally, given a basis that minimizes the spread of the wavefunction, there are different choices for quantifying the spread. A natural chouce is the average position along the chain
\be 
C(t) = C_\mathcal{K}(t) = \sum_{n} n \,\vert \braket{K_n}{\psi(t)}\vert^2 =
\sum_{n} n \,p_n(t)\;.\label{eq:spread_comp_def}
\ee
A related notion is the effective dimension of the Hilbert space explored by the time evolution, namely
\be 
C_{{\rm dim}}(t)
= e^{H_\textrm{Shannon}}=e^{-\sum\limits_n p_n(t)\log p_n(t)}\;.
\label{ecom2}
\ee
$C(t)$ and $C_{{\rm dim}}(t)$ have different behavior in chaotic and integrable systems (see \cite{SpreadC,Balasubramanian:2023kwd,Balasubramanian:2024ghv}).\footnote{An intriguiging class of ``pseudo-integrable'' systems shows behavior that is intermediate between chaos and integrablity, and would be interesting to investigate further using these tools \cite{Balasubramanian:2024ghv}.}
Finally recall that, as described above, the wavefunction in the Krylov basis is a functional of the Lanczos coefficients, and that these are in turn determined  by the survival amplitude.  One consequence will be that, in what follows, saturation of the survival amplitude will also imply saturation of the spread complexity.

\section{Saturation of the Einstein-Rosen bridge dual to DSSYK}\label{III}

In this section we will connect the SYK model and its doubled scaled limit, random matrix theory, and spread complexity, in a way that allows us to understand why the Einstein-Rosen bridge dual to the large $N$ SYK model saturates in size at late times.

\subsection{Hilbert spaces from Thermofield doubles}

Eternal black holes in AdS space are dual to Thermofield Double (TFD) states of a product of two identical CFTs with Hilbert spaces $\mathcal{H}_L$ and $\mathcal{H}_R$ and Hamiltonians $H_L$ and $H_R$.  To define a TFD state in terms of energy eigenstates $\vert n\rangle$ and eigenvalues $E_n$ of the two CFTs we construct
\be 
\vert\psi_{\beta}\rangle \equiv\frac{1}{\sqrt{Z_{\beta}}}\sum_n e^{-\frac{\beta E_n}{2}}\vert n,n\rangle\;.
\label{TFDdef}
\ee
The TFD is invariant under evolution with Hamiltonian $H_L-H_R$, where $H_{L,R} = H$ act independently on ${\cal H}_{L,R}$. This invariance is related in the gravitational dual to the isometries of the eternal black hole associated with the Schwarzschild  time coordinate. This means that maximal volume slices do not grow under evolution with $H_L-H_R$, i.e.,  slices between two times $t_L=t$ and $t_R=-t$ have the same volume for all $t$. 

However, the TFD state is not invariant under evolution by the action of a single Hamiltonian, say $H_L\equiv H$, or equivalently by the action of the ``global time'' evolution $(H_L + H_R)/2$. Concretely, unitary evolution with a single Hamiltonian gives
\be 
\vert\psi_{\beta} (t)\rangle=e^{-iHt}\vert\psi_{\beta}\rangle=\vert\psi_{\beta+2it}\rangle=\sum_n e^{-(\beta+2it) E_n/2}\,\vert n,n\rangle\;.
\ee
For a chaotic theory this evolution should, in some sense, make the state more ``complex''. The proposal in \cite{Susskind:2014rva,Stanford:2014jda} is that this complexity should be quantified in the gravitational descriptoion by the volume of maximal volume slices between say $t_L=0$ and $t_R=t$. These slices go through the interior of the black hole, i.e., they thread the ER bridge, and, after a short initial period, grow linearly in time forever in the semiclassical limit. However, any reasonable measure of complexity will be bounded in a system with a finite Hilbert space. So, as explained in \cite{Balasubramanian:2022gmo,Balasubramanian:2022lnw,Iliesiu:2024cnh}, if complexity and ER bridge volume are related, the bridge size should eventually saturate.  This saturation is not visible in perturbative gravity and requires the inclusion of non-perturbative corrections to the gravitational path integral.

How can we construct the required finite dimensional Hilbert space for the eternal black hole?  Following \cite{SpreadC}, one option is to construct the Krylov basis associated to the TFD state. As reviewed in the previous section, such a basis is fully determined by the moments of the Hamiltonian in the initial state. In the TFD state these moments read
\be\label{Momb}
M_n \equiv \langle\psi_{\beta}\vert\, H^n \,\vert\psi_{\beta} \rangle
=\frac{1}{Z_\beta}\,\textrm{Tr}\left( e^{-\beta H}  H^n\right) \;,
\ee
i.e., they are the  moments of the Hamiltonian in a thermal state. Equivalently, the Krylov basis for this evolution is fully determined by the survival amplitude, which here reads
\be \label{ACPF}
\langle\psi_\beta\vert\psi_{\beta} (t)\rangle= Z_{\beta-it}\;.
\ee
i.e., it is the analytically continued partition function of the theory.  The spectral form factor, defined as
\be \label{SFF}
S(t)=\vert Z_{\beta-it}\vert^2 \;,
\ee
is then the survival probability of a dynamical process, corresponding to time evolution of the TFD state.\footnote{This last fact has been noticed and used for different purposes in \cite{Papadodimas:2015xma,delCampo:2017bzr,Verlinde:2021jwu,Stanford:2022fdt}.} All of these quantities  have been extensively studied in random matrix theory and quantum gravity \cite{Guhr_1998,Cotler:2016fpe}. Moreover, the partition function and the energy moments have simple geometric duals. In  2d gravity they can even be defined non-perturbatively \cite{Saad:2019lba,Stanford:2019vob,Johnson:2019eik,Maxfield_2021,Mertens_2021,Witten:2020wvy}. 

So from the survival amplitude~(\ref{ACPF}) or the Hamiltonian moments~(\ref{Momb}), we can construct the Krylov basis
\be 
\vert 0\rangle\;,\,\vert 1\rangle\;,\,\vert 2\rangle\;,\,\cdots\;,\,\vert L\rangle\;,
\ee
just by iterating the Lanczos algorithm. In this Krylov basis the initial vector is the TFD, $\vert 0\rangle =\vert\psi_\beta\rangle$, while the last vector $\vert L\rangle$ sets the dimension of the explored Hilbert space (see \cite{SpreadC} for details).  For the SYK model with any fixed number of fermions, this Hilbert space will be finite.\footnote{In a later section we will consider similar constructions for higher dimensional theories with infinite dimensional Hilbert spaces.  In that case the TFD state will have support on the entire infinite dimensional energy eigenbasis.  So to construct the finite dimensional Hilbert space associated to states within a definite energy band, it will be more useful to consider microcanonical TFD's, or more generally a TFD in which we include a further weight function that selects a particular energy range. 
}
Next, note that, as  reviewed above, the Hamiltonian driving the evolution takes a tridiagonal form $T$ in the Krylov basis, with diagonal elements $a_n$ and off-diagonal elements $b_n$. These can be computed from~(\ref{ACPF}). Then, in the Krylov basis we can write the moments of the Hamiltonian as
\be \label{momentst}
M_n =\frac{1}{Z_\beta}\,\textrm{Tr}\left( e^{-\beta H} H^n\right)=\langle \psi_{\beta}\vert\, T^n\,\vert \psi_{\beta}\rangle \;,
\ee
namely as a transition amplitude from the TFD and back, with a transition matrix given by the tridiagonal version of the Hamiltonian.

This then is a canonical way to construct the finite dimensional Hilbert space explored by time evolution from the TFD state and its ER bridge dual.
Note that in the strict double scaling limit of the SYK model and the related variants of semiclassical JT gravity that appear as effective descriptions at low energies, this Hilbert space turns out to be infinite dimensional.   In the DSSYK model this is because the theory in the scaling limit has an infinite number of degrees of freedom.  On the JT side this is because the phase space is characterized by a continuous parameter, the length of ER bridges \cite{Harlow:2018tqv}.  But in the quantum theory with finite coupling, finite number of degrees, and $\hbar>0$ the underlying Hilbert space is discrete and finite dimensional even if this is hard to see in the low-energy semiclassical limit that is well described by JT gravity, and by the double scaled limit.  We will see how to maintain this underlying discreteness and finiteness, which becomes important for understanding the very late time dynamics, even as we approach the thermodynamic/classical limit of the theory.

Notice the DSSYK tridiagonal method \cite{Berkooz_2018} to find the moments, Eq.~(\ref{trimo}), is a special case of the construction outlined above. It corresponds to considering the Krylov basis for TFD state of the DSSYK model with $\beta \to 0$. In this case the initial state is the infinite temperature TFD of the microscopic SYK theory, and the chord Hilbert space is precisely the Hilbert space spanned by its time evolution. It is not just an auxiliary construction. The chord basis is exactly the Krylov basis. But since the microscopic Hilbert space is finite dimensional, if we are not strictly in the thermodynamic limit we should be able to find a finite dimensional tridiagonal transition matrix $T$ providing the same moments. Equivalently, the matrix $T$ in Sec.~(\ref{trimo}) from \cite{Berkooz_2018} should be understood as an approximation that should be corrected so that the off-diagonal elements go to zero as we approach the end of the matrix. We turn to this problem now.

\subsection{Lanczos spectrum of DSSYK: numerical results}
We now want to find the Lanczos coefficients corresponding to DSSYK given a TFD initial state with temperature $\beta$. In principle, with enough computational power, the Lanczos coefficients could be obtained by following the steps in \cite{SpreadC}. To do so, we would produce  instances of the SYK Hamiltonian, diagonalize them, construct the TFD at the appropriate temperature, and finally find the Hessenberg form of the Hamiltonian matrix. The last step has to be done in a basis in which the first state is the TFD. While one can do this for SYK with a limited number of fermions \cite{SpreadC}, this procedure is out of reach for the large systems required to approach the double-scaling limit.

\begin{figure}
    \includegraphics[width=0.98\linewidth]{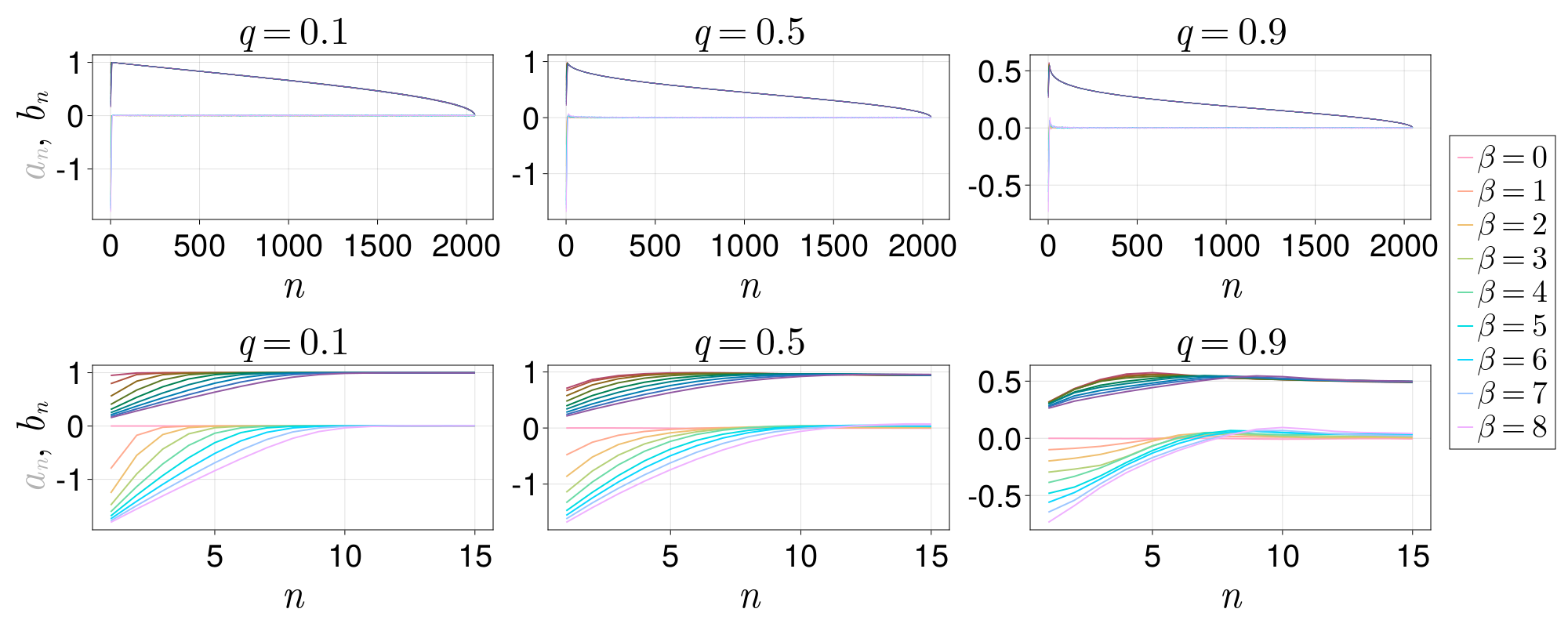}
    \caption{\label{numerical_LZ} The Lanczos spectrum ($a_n$ in lighter colors and $b_n$ in darker colors) for DSSYK at different values of $q=e^{-\lambda}$ given in the titles, corresponding to the evolution of the TFD at different temperatures $\beta$ given in the legend. Only the edge near $n=0$ differs between the different temperature initial states.
    This is a numerical computation using the leading density of states. It is transparent that the bulk of the Lanczos spectrum, namely $n\sim \mathcal{O}(L)$ that corresponds to the descent, does not depend on the temperature. It can then be approached with the RMT techniques described in the text.     }
\end{figure}

To overcome this difficulty we use the results of \cite{Balasubramanian:2022dnj,Balasubramanian:2023kwd}. There it was shown that in the thermodynamic limit, the average Lanczos spectrum only depends on the leading density of states. The average here is taken over the SYK ensemble, or, if we want to consider a particular member of the ensemble in the thermodynamic limit, it is a coarse-grained, locally averaged, continuum density of states.\footnote{As shown in \cite{Balasubramanian:2023kwd,Balasubramanian:2022dnj}, while the detailed Lanczos spectrum of individual SYK instances does depends on correlations, the average, taken in the senses described in the text, does not. The correlations in the energy spectrum, which match those of a random matrix theory, manifest themselves in the two point correlations of the Lanczos spectrum, not in the one point functions.}   For DSSYK, the density of states $\rho(E)$ is  (\ref{rd}). This can be written in terms of a Jacobi $\theta_1$ function as
\be\label{dsde}
   \rho(E)= \rho(2\cos(\theta))=\frac{\theta_1(z,q^{1/2})}{2\pi q^{1/8}}\;.
\ee
Here we are expressing energies in units of $\mathcal{J}$ -- the normalization is explained below Eq.~\ref{eq:SYKcouplings}.
We can now emulate to very good precision the DSSYK scenario by finding the Hessenberg form of a matrix with a GUE spectrum stretched to match $\rho(E)$. This stretching is achieved by applying the cumulative distribution function, see \cite{Balasubramanian:2023kwd} for a more detailed explanation of this method and further applications. The results are plotted in Fig.~\ref{numerical_LZ}. 

Fig.~\ref{numerical_LZ} shows the Lanczos coefficients for different values of $q$ and $\beta$. The earliest part of the spectrum for $\beta = 0$ reproduces Eq.~\ref{transf} that was obtained by the chord Hilbert space methods of \cite{Berkooz_2018}, where the authors are effectively working at infinite temperature, and with an infinite dimensional Hilbert space to find their approximate solution via the double scaling limit to the large $N$ SYK model.    Additionally, we see from Fig.~\ref{numerical_LZ} that for larger $n$ there is a universal structure for all $\beta$.  Specifically, $a_n$ is essentially zero and $b_n$, the hopping amplitude on the Krylov chain, declines steadily to zero in exactly the same way for all $\beta$ and fixed $q$.  Here we are considering the Krylov chain starting with the thermofield double state, but the Laczos coefficient descent is actually the same for the initial random states considered in \cite{Balasubramanian:2022dnj,Balasubramanian:2023kwd}. The descent is required to achieve a finite tridiagonal Hamiltonian transition matrix appropriate to a finite dimensional Hilbert space. We will analyze this Lanczos spectrum analytically in the next section.

From the present analysis, the infinite transition matrix derived in \cite{Berkooz_2018}, Eq.~(\ref{transf}), can be regarded as an approximate description, extrapolated from the initial part $O(1)$ part of this matrix when  $L \to \infty$. Our finite matrix is the actual Hamiltonian of the theory, with size equal to the dimension of the Hilbert space, and reproduces the right moments (\ref{moman}) and density of states (\ref{rd}) for large but finite SYK models.  We arrived at these results via the moments in (\ref{momentst}). This means, as discussed in Sec.~\ref{sec:sykReview}, that the Krylov basis that we are producing is precisely the same as chord Hilbert space of \cite{Berkooz_2018}, but adapted to finite and large $N$ and without needing the strict $N\to \infty$ limit.  By construction, precisely this finite dimensional Hilbert space is explored by the time evolution of the TFD in the microscopic SYK description.  Thus we see that the chord Hilbert space of \cite{Berkooz_2018} should be understood as physical, not just an auxiliary construction, but approximate as a description of the quantum theory away from the large $N$, semiclassical limit. At finite $N$, the right Hilbert space and Hamiltonian is the one presented above. We will see later how this predicts the saturation of the Einstein-Rosen bridge.

\subsection{Lanczos spectrum of DSSYK: analytical results}
\label{sec:analytical}

We will now derive the  Lanczos spectrum analytically. In particular we want to derive the descent to zero of the hopping coefficients $b_n$, which is the key finite size correction to the results of \cite{Berkooz_2018}.\footnote{We will separately treat the first $n\sim O(1)$ order Lanczos coefficients $a_n, b_n$ which can be approximated via the moments of the Hamiltonian in the thermal ensemble $M_n=\frac{1}{Z_\beta}\int_{-2}^2 E^n e^{-\beta E}\,\rho(E) dE $
and then using the recursion method \cite{viswanath2008recursion, SpreadC}.} There are two equivalent ways to do this. These were described in \cite{Balasubramanian:2022dnj} for random matrix theories, but it was later noticed in \cite{Balasubramanian:2023kwd} that coarse-grained distribution for $a_n,b_n$ can be derived in the thermodynamic limit for any system for which we know the density of states.

The first method uses a continuity assumption for the Lanczos coefficients in the thermodynamic limit \cite{Balasubramanian:2022dnj} which says that the Lanczos coefficients $a_n,b_n$, when expressed in terms of the variable $x\equiv n/L$, become continuous functions in the large-$L$ limit.\footnote{See \cite{doi:10.1063/1.533010,PhysRevD.47.1640} for further considerations about this continuity assumption.} To have such a limit, 
the Lanczos coefficients must vary sufficiently slowly within any block of size $l$ which satisfies $l/L \to 0$ as $L\to\infty$. We can then approximate the Hamiltonian as a direct sum of such blocks within which the  Lanczos spectrum is constant. A constant tridiagonal matrix can be diagonalized easily, and its density of states can be found as well. This leads to a consistency equation, namely the total density of states must  equal  the sum of density of states of each block at position $x$. This is encapsulated in  integral equation
\be \label{intdl}
\rho(E) = \int_0^1 dx\, \frac{\Theta(4\,b(x)^2-(E-a(x))^2)}{\pi\, \sqrt{4\,b(x)^2-(E-a(x))^2}}\;.
\ee
In this equation $\Theta(x)$ is the Heaviside step function, and $a(x)\approx a_{nL}, b(x)\approx b_{nL}$ are the large $L$ continuous approximations of the Lanczos coefficients. Starting from the density of states, which we assume as given, this equation provides the leading Lanczos spectrum at large $L$. From this continuous spectrum we obtain a finite tridiagonal matrix, namely one with precisely $a_{n}\equiv a(x)$ and $b_{n}\equiv b(x)$, for $x= n/L$. By construction, in the thermodynamic limit, this finite tridiagonal matrix gives precisely the right moments in (\ref{momentst}). This method can be applied Hamiltonians with any density of states \cite{Balasubramanian:2023kwd}. In particular it applies to the DSSYK formula in (\ref{dsde}).

The second method starts from the observation  in \cite{Balasubramanian:2022dnj,Balasubramanian:2023kwd} that for any theory in the thermodynamic limit the Lanczos spectrum  locally averaged over the Krylov chain  only depends on the density of states, as it also does for a random matrix theory and  SYK ensembles.  This means that the average SYK Lanczos spectrum that we seek will be the same as that for a random matrix theory with the same density of states.  To define the latter, we have to select a potential $V(E)$ leading to the desired density of states via (\ref{sadrh}).  As shown in \cite{Balasubramanian:2022dnj,Balasubramanian:2023kwd} a change of variables for an RMT defined in such a way gives the joint probability distribution for the Lanczos spectrum.   In the thermodynamic limit this distribution becomes sharply peaked, and so we can derive the Lanczos spectrum from its saddlepoints -- in effect the logarithm of the distribution acts as an effective action whose equation of motion gives the Lanczos spectrum \cite{Balasubramanian:2022dnj}:
\bea\label{sadlan}
    4(1-x) &=& b(x) \frac{\dx}{\dx b(x)}\gr{\int dE \frac{V(E)}{\pi \sqrt{4b(x)^2 - (E-a(x))^2}}}\;,\nonumber\\
    0&=&\frac{\dx}{\dx a(x)}\gr{\int dE \frac{V(E)}{\pi \sqrt{4b(x)^2 - (E-a(x))^2}}}\;.\label{eq:one_point_final}
\eea
This method is consistent with the previous one but more powerful since it can be extended to obtain covariances of the Lanczos coefficients. However, it requires  computation of the potential $V(E)$ from the density of states using (\ref{sadrh}).

We now work out the potential associated with the DSSYK density of states. To this end we first expand the Jacobi $\theta_{1}$ function in (\ref{dsde}) as 
\be
\rho(2\cos(\theta))=-\frac{\theta_1(\theta,q^{1/2})}{2\pi q^{1/8}}=\frac{1}{\pi}\sum_{n=0}^\infty (-1)^nq^{n(n+1)/2}\sin((2n+1)\theta)\;.
\ee
We now define scaled Chebyshev polynomials $C_n (x)$ by
\be
C_0(x)=1 ,\,\,\,\,\,\,\, C_1(x)=x ,\,\,\,\,\,\,\, C_i=xC_{i-1}-C_{i-2}\;.
\ee
The explicit solution is $C_n(2\cos \theta)=\frac{\sin((n+1)\theta)}{\sin(\theta)}$. In terms of these polynomials the density of states is then
\be
\rho(E)=\frac{1}{2\pi}\sum_n (-1)^nq^{n(n+1)/2}C_{2n}(E)\sqrt{4-E^2}\;.\label{dssyk_density_of_states}
\ee
To find the potential it is helpful to define another auxilliary set of polynomials:
\be
\label{eq:pn}
P_n(\omega)\equiv \text{p.v.}\int dE~\frac{C_n(E)\sqrt{4-E^2}}{E-\omega}\;.
\ee
We now now notice that
\be 
\text{p.v.}\int dE~(E-\omega)\, C_n(E)\,\frac{\sqrt{4-E^2}}{E-\omega}=\int dE~C_n(E)\,\sqrt{4-E^2}=0\;,
\ee
since $C_0=1$ and $C_n$ are orthogonal polynomials. This implies that
\bea
P_n(\omega)&=& \text{p.v.}\int dE~\frac{C_n(E)\sqrt{4-E^2}}{E-\omega}=\text{p.v.}\int dE~(EC_{n-1}(E)-C_{n-2}(E))\frac{\sqrt{4-E^2}}{E-\omega}\nonumber\\
&=&\text{p.v.}\int dE~(\omega C_{n-1}(E)-C_{n-2}(E))\frac{\sqrt{4-E^2}}{E-\omega}=\omega P_{n-1}(\omega)-P_{n-2}(\omega)\;.
\eea
Direct calculation shows that $P_0(\omega)=3\omega$, and $P_1(\omega)=3\omega^2-6$. These polynomials can be expanded in a basis of the scaled Chebyshev polynomials $C_n(2\cos \theta)=\frac{\sin((n+1)\theta)}{\sin(\theta)}$ and $T_n(2\cos \theta)=\cos(n\theta)$ where $T_0(x)=1$, $T_1(x)=x/2$ and $T_n=xT_{n-1}-T_{n-2}$. Explicitly
\be
\label{eq:pnTOcn}
    P_n(\omega)=3\omega C_n-12\frac{C_n-T_n}{\omega}.
\ee   
Inserting now the density of states~(\ref{dssyk_density_of_states}) into the saddle point equation~(\ref{sadrh}), we can write $V'(\omega)$ in terms of the second Jacobi theta function. We first write
\be
\frac{1}{4}V'(\omega)=\frac{1}{2\pi}\sum_n (-1)^nq^{n(n+1)/2}P_{2n}(\omega)
    =\frac{3}{2\pi}\sum_n (-1)^nq^{n(n+1)/2}\left(\omega C_{2n}-4\frac{C_{2n}-T_{2n}}{\omega}\right)\nonumber\;,
\ee
where we used (\ref{eq:pn}) and (\ref{eq:pnTOcn}).   So: 
\bea\label{dssyk_RMT_potential}
    V'(2\cos\theta)&=& \frac{6}{\pi}\sum_n (-1)^nq^{n(n+1)/2}\left(\frac{2\cos\theta \sin((2n+1)\theta)}{\sin(\theta)}-4\frac{\sin((2n+1)\theta)-\cos(2n\theta)\sin(\theta)}{2\cos\theta \sin\theta}\right)\nonumber\\
    &=& \frac{12}{\pi}\sum_n (-1)^nq^{n(n+1)/2}\frac{\cos\theta \sin((2n+1)\theta)-\sin(2n\theta)}{\sin\theta}\nonumber\\
        &=& \frac{12}{\pi}\sum_n (-1)^nq^{n(n+1)/2}\cos((2n+1)\theta)\;.
\eea

\begin{figure}
    \includegraphics[width=0.98\linewidth]{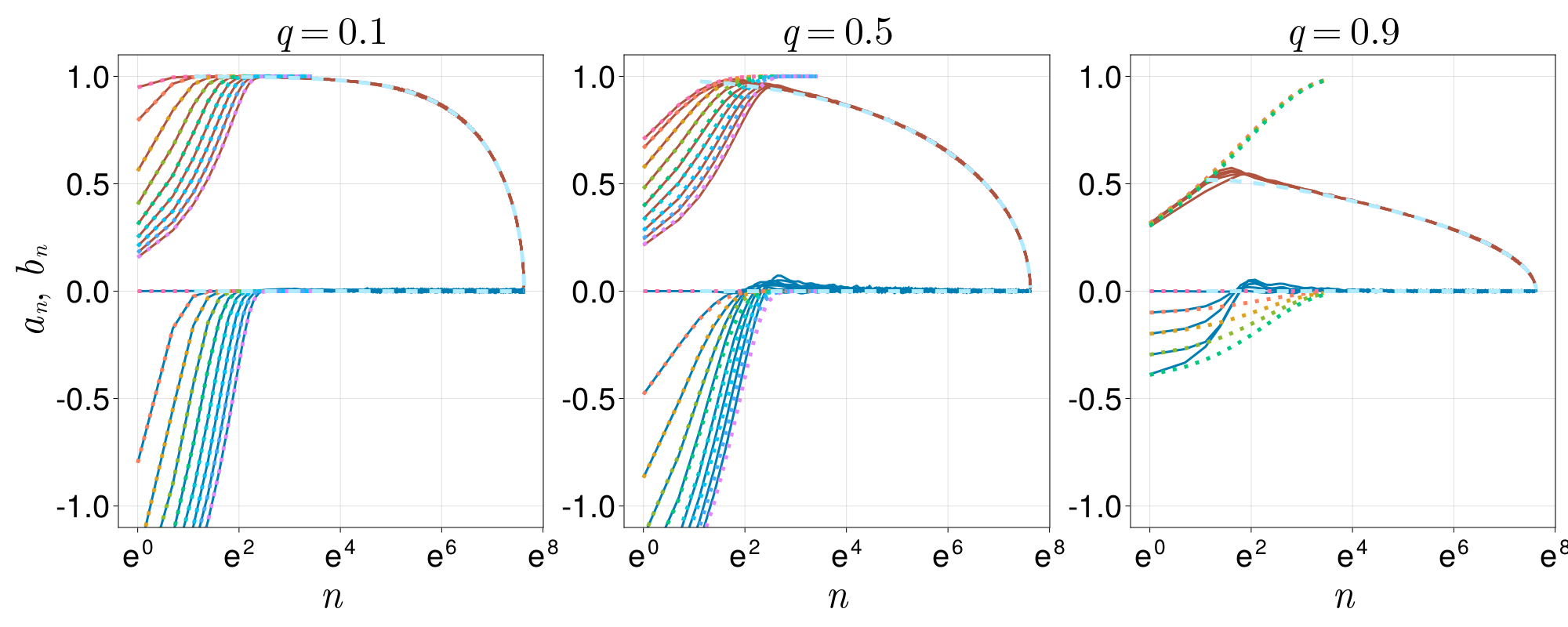}
    \caption{\label{numerical_and_analytical} The leading (large-L) Lanczos spectrum for DSSYK computed analytically using the integral equation~(\ref{intdl}) or the saddle point equations \eqref{dssyk_RMT_potential}, valid for $n\gg 1$ (but still $n\sim\mathcal{O}(1)$ in the large-$L$ limit) (dashed lines). We also include the small $n\sim O(1)$ analytical values at various $\beta$, corresponding to the same colors as in Fig \ref{numerical_LZ} (dotted lines). We also show the numerical results for comparison (solid lines). The $x$-axis is scaled logarithmically to appreciate the initial (growing) part of the spectrum. Notice the dependence on the temperature.}
\end{figure}

Either inserting \eqref{dssyk_density_of_states} in the integral equation~(\ref{intdl}) or inserting the potential \eqref{dssyk_RMT_potential} into the saddle point equations~(\ref{sadlan}) we finally arrive at the thermodynamic limit of the average Lanczos spectrum. We were not able to solve these equations in closed form. But in Fig.~\ref{numerical_and_analytical} we plot the numerical solutions and find perfect agreement with the numerical evaluations of the previous section.\footnote{Note that our analytical solutions are valid for $L\gg n\gg 1$ in the large-$L$ limit), i.e., they do not capture the initial growing part computed in \cite{Berkooz_2018}. This growing part can be computed by the moment method as discussed earlier. Still, our analytical solution for the bulk of the Lanczos spectrum covers most of the Lanczos coefficients.}

We finish this section with an aside that is not needed in the rest of the article. As explained in \cite{Balasubramanian:2022dnj,Balasubramanian:2023kwd}, the RMT approach also gives us also the covariances of the Lanczos coefficients. In these references the covariances were only computed for a random initial state, namely a state created by a applying a Haar random unitary to a given reference state in the Hilbert space. Unlike the mean Lanczos spectrum which will be identical to that of the original theory in the thermodynamic limit, the covariances will only match the original if it too has RMT correlations.  Fortunately, as discussed earlier, the SYK model does have RMT statistics for its correlations.  So we can proceed with our method.

\begin{figure}
    \includegraphics[width=0.98\linewidth]{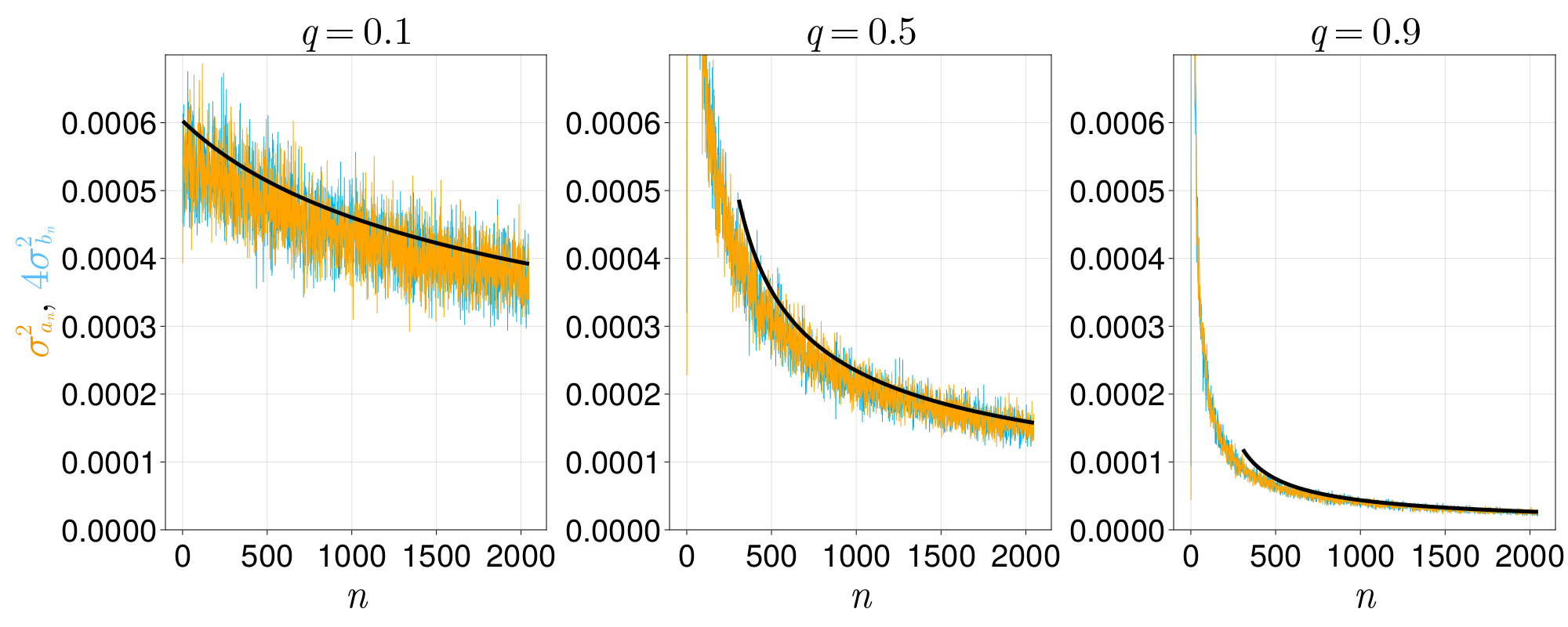}
    \caption{\label{variance} The variance of the (large-L) Lanczos spectrum  for DSSYK with random initial states computed analytically using \eqref{cov} with a polynomial approximation, valid for $n\gg 1$, plotted in solid black. We also show numerics (orange for $\sigma_{a_n}^2$ and blue for $4\sigma_{b_n}^2$) for comparison.
    }
    \label{fig:variances}
\end{figure}

For random initial states, by examining fluctuations around the large-L saddlepoint of the distribution for the Lanczos spectrum, i.e., perturbations around solutions of the saddle point equations~(\ref{sadlan}), the authors of \cite{Balasubramanian:2022dnj} arrived at an effective Gaussian system determining the covariances. The final formulas are
\bea\label{cov}
\text{cov} (a_i,a_{i+\delta}) \approx 4\,\text{cov} (b_i,b_{i+\delta}) \approx \frac{1}{2\pi\beta_D L} \int_0^{2\pi}\frac{e^{2ik\delta}dk}{\lambda(k, a_i, b_i)}\;,\nonumber\\
2\,\text{cov} (a_i,b_{i+\delta}) \approx \frac{1}{2\pi\beta_D L} \int_0^{2\pi}\frac{e^{ik(2\delta-1)}dk}{\lambda(k, a_i, b_i)}\;,
\eea
where we define
\bea\label{lrmt2}
    \lambda(k, a_i, b_i) =\int dE~\frac{V'(E)-V'(a_i)}{b_i(E-a_i)}\,\eta\gr{\frac{E-a_i}{b_i},e^{ik}}\;,\label{eq:eig_defs}
\eea
together with 
\bea
    \eta(x,t)=\frac{x}{\pi\sqrt{4-x^2}}\frac{1}{t+\frac{1}{t}-x}\;.
\eea
In Fig~\ref{fig:variances} we plot the variances of the Lanczos spectrum and compare with the numerical evaluations of DSSYK. The match is fairly good. We attribute the small discrepancy to the fact that the continuity assumption in \cite{Balasubramanian:2022dnj} takes longer to converge for higher order polynomial terms in the random matrix potential, as well as the finite truncation of the Cheybyshev polynomials. Still, we see numerically that the variance of the $a$'s is equal to four times the variance of the $b$'s, a prediction from Ref. \cite{Balasubramanian:2022dnj}. 

\subsection{From classical growth of the ER bridge to saturation}

\begin{figure}
    \includegraphics[width=0.98\linewidth]{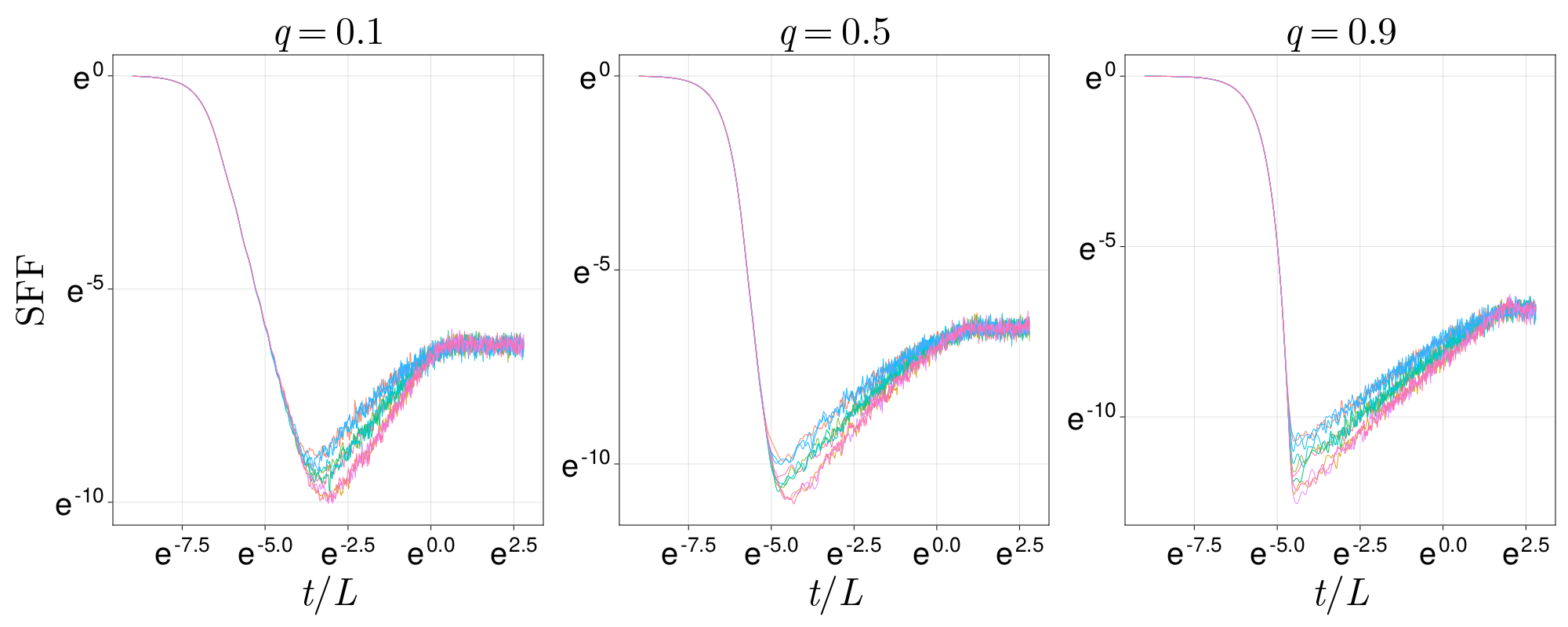}
    \caption{\label{AZ_SFF} The spectral form factor for DSSYK associated with the TFD state at temperature $T=1$ and different values of $q=e^{-\lambda}$. Different colors correspond to different quantum chaotic universality classes. We see that at a qualitative level, the spectral form factor only depends on the Dyson index $\beta$, and not on the parameter $\alpha$ characterizing the Altland-Zirnbauer class.}
\end{figure}

\begin{figure}
    \includegraphics[width=0.98\linewidth]{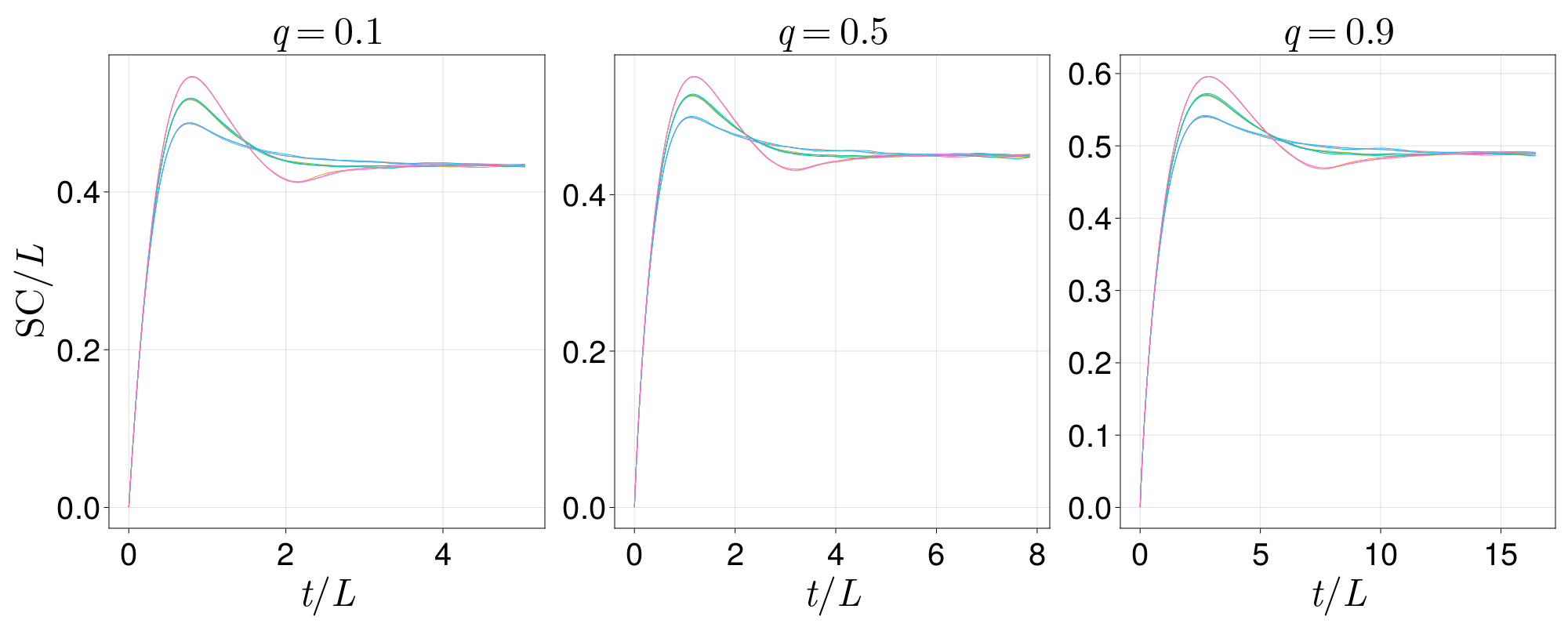}
    \caption{\label{AZ_SC} The spread complexity for DSSYK associated with the TFD state at temperature $T=1$ and different values of $q=e^{-\lambda}$. Different colors correspond to different quantum chaotic universality classes. We see that at a qualitative level, the spread complexity (as the spectral form factor) only depends on the Dyson index $\beta$, and not on the parameter $\alpha$ characterizing the Altland-Zirnbauer class.}
\end{figure}

Recall that in the triple scaling limit the SYK model has a continuum effective description \cite{Berkooz_2018} in which the Hamiltonian on the discrete chord Hilbert space becomes equal to the continuum Liouville Hamiltonian of JT gravity \cite{Harlow:2018tqv}. Also,  \cite{Rabinovici:2023yex} showed that in this  limit the spread complexity in the infinite dimensional chord basis is identically equal to size of the JT gravity Einstein-Rosen bridge.  The latter authors found that this spread complexity shows the linear growth in time required to reproduce the growing size of classical ER bridge at sub-exponential times.  However, because they were working in the large $N$ limit with a continuum description which is effectively classical, the growth continues forever.   The challenge is to understand how to back away from this limit. In other words, we want to compute the non-perturbative quantum effects that correct the classical, continuum picture, and reveal the underlying discreteness and finiteness of the Hilbert space at times exponentially large in the entropy. But we seek to achieve this while nevertheless retaining the continuous geometric effective description at early times. 

The authors of \cite{Iliesiu:2024cnh} approached this problem in continuum JT gravity by inferring an inner product on the continuum states of fixed ER bridge size that is modified by non-perturbative gravitational effects.  This modified inner product renders the Hilbert space finite and leads to late time saturation, by a mechanism similar to the black hole microstate counting in \cite{Balasubramanian:2022gmo,Balasubramanian:2022lnw}.  Here we take a different complementary approach.  Instead of using the energy basis to induce a discretization of the Hilbert space in the length basis as in \cite{Iliesiu:2024cnh}, we will provide a fundamental discretization of the length basis directly. To this end, we start with an underlying discrete SYK model, and recognize that the continuum limit  giving rise to JT gravity in \cite{Berkooz_2018} is an extrapolation from an exponentially small part of the full Hamiltonian. Namely, in the tridiagonalized language, the authors of \cite{Berkooz_2018}  extrapolate from Lanczos coefficients $a_n$ and $b_n$ with $n/L \to 0$ as $L$ becomes large.  Noting that $L \sim e^S$ where $S$ is the entropy of the system, we see that the continuum limit is an extrapolation from an exponentially small part of the Hilbert space which is ``blown up'' to infinite size by the large $N$ limit.   Our analysis in Sec.~\ref{sec:analytical} above showed that for $n\sim \mathcal{O}(L)$ the Lanczos $b_n$ coefficients cease to be constant and descend to zero. The rate of change is $\mathcal{O}(1/L)$, i.e., it is non-perturbative from a gravitational perspective because it is $\mathcal{O}(e^{-S})$. From the JT gravity point of view the associated effects arise from  wormholes in the gravitational path integral.  But from our point of view, we can precisely compute the requisite effects directly from analysis of the underlying Hamiltonian, and then take the large $N$ limit while keeping both the $\mathcal{O}(1)$ part that leads to the classical continuum gravity description at early times, and the $\mathcal{O}(L)$ parts that dominate the quantum effects at late times.   The latter show the non-perturbative Lanzos descent  controlled by the saddle point equations (\ref{sadlan}), and imply saturation of spread complexity.  If we take the spread complexity to be the non-perturbative definition of the ER bridge size, matching the standard description at early times, then the ER bridge will saturate at late times.

In this section we will analyze the saturation. Since  spread complexity is a functional of the analytically continued partition function, we will also be able to analyze the spectral form factor of the theory. Direct numerical computation starting with SYK in its double scaled limit is out of reach. There are two methods to approach this problem. The first  approximates the Hamiltonian of the theory through the
analytical formulas for the one and two-point functions of the Lanczos spectrum discussed above -- see \cite{Balasubramanian:2022dnj} for  detailed derivations. In this case, since the Hamiltonian is tridiagonal, we only have to work with $\mathcal{O}(L)$ matrix elements, rather that $\mathcal{O}(L^2)$.  With this approximation of $H_{\textrm{trid}}^{\textrm{approx}}$ in hand, we can compute the spectral form factor simply from
\be
    Z_\beta = \textrm{Tr}(e^{-\beta H})=\textrm{Tr}(e^{-\beta H_{\textrm{trid}}})\approx \textrm{Tr}(e^{-\beta H_{\textrm{trid}}^{\textrm{approx}}}),
\ee
and the spread complexity following from this \cite{SpreadC}.

In the second method we recall, as noted in Sec.~\ref{sec:SYK-RMT}, that large $N$ SYK models and variants of JT gravity, have chaotic Hamiltonians that realize the correlations of different random matrix classes.  This means that we can model the full large $N$ SYK theories by constructing random matrix theories in different unversality classes, and with potentials tuned to reproduce the desired SYK density of states.  Random matrix Hamiltonians with general potentials can be generated  via Metropolis sampling \cite{hastings1970monte}, but this is a slow procedure. A better method is to stretch the spectra of Gaussian random matrix theories to match the density of states of the target non-Gaussian theory. To this end we apply to each energy eigenvalue the cumulative distribution $f(E)=\frac{1}{2\pi}\int_{-2}^{E} dE'~\sqrt{4-E'^2}$ associated with the Gaussian RMT, followed by the inverse cumulative distribution $g(E)=\int_{-\infty}^E dE'~\rho(E')$ of the density of states of the target non-Gaussian RMT. This leads to an approximation $\tilde{E}_i=g^{-1}(f(E_i))$ of the spectrum of the non-Gaussian Hamiltonian, see \cite{Balasubramanian:2022dnj,Balasubramanian:2023kwd} for more details. Both methods provide very accurate results.

In Figs.~\ref{AZ_SFF} and \ref{AZ_SC}  we use the second method to compute the spectral form factor and the spread complexity for temperature $\beta=1$ and different values of the SYK parameter $q$.  We also examine the three Dyson ensembles and the seven Altland-Zirnbauer classes that share the same desnity of states as the DSSYK model. We find a strong dependence of the Dyson $\beta$ index, and mild dependence on the Altland-Zirnbauer $\alpha$ index. In  these scenarios, all of which describe chaotic theories, the spread complexity reaches a peak and then displays a downward slope before saturation, as shown in \cite{SpreadC}. This peak and slope would not have appeared if we had take an spectrum with the same density of states but  Poisson distributed eigenvalues as expected for integrable theories \cite{SpreadC}.

We now adopt the identification of spread complexity with ER bridge size in  \cite{Rabinovici:2023yex}.  At sub-exponential times in the evolution of the TFD state our computation will match \cite{Rabinovici:2023yex} precisely because we are computing the Lanczos coefficients through precisely the same moment method used in \cite{Berkooz_2018} and \cite{Rabinovici:2023yex}.  Thus the spread will be identical and will reproduce the classical ER bridge size growth as seen in those works.  However, as discussed above, those authors worked in a limit where the Hamiltonian is infinite dimensional, effectively blowing up the sub-exponential part of the dynamics to infinite size.  If we instead work with the actual large, but finite dimensional system, then at exponential times the finiteness of the Hilbert space leads to a cross-over to a regime where the ER bridge size, still identified with the spread complexity, {\it decreases} and then saturates at a plateau. This is one of the main conclusions of this article. If we non-perturbatively define the Einstein-Rosen bridge volume through the Krylov basis, which matches the chord basis of \cite{Berkooz_2018} for sub-exponential basis elements, then the dynamics of ER bridges saturates at exponentially late times in a manner appropriate to the underlying quantum chaotic universality class. In particular the saturation follows the ramp-peak-slope-plateau structure described in \cite{SpreadC}.\footnote{For Dyson index $\beta=4$, corresponding to the symplectic universality class, we have a slope and then an upward ramp again to the plateay. This is the same behavior seen for the spectral form factor in that universality class. It is related to the coherence of the wavefunction in the Krylov basis, see \cite{Balasubramanian:2023kwd} for a more detailed explanation and for the analysis of more general Dyson $\beta$ indices.} We were not able to find closed form formulae for the complete time evolution,\footnote{This would require the solution of a one dimensional hopping chain with Gaussian random local couplings, whose one and two point functions follow the saddle point equations described above.} but we were able to calculate the saturation plateau analytically. We turn to this next.

\subsection{Saturation value}

The authors of \cite{Susskind:2014rva,Stanford:2014jda} conjectured that the volumes of ER bridges measure some notion of complexity of the underlying quantum state. For a black hole of mass $M$, this  suggests that ER bridge sizes should saturate at $\mathcal{O}(e^S)$, where $S$ is the black hole entropy at mass $M$.  We will now compute the saturation value if we extrapolate the identification ``bridge size = spread complexity'' valid at sub-exponential times, to late times that are exponential in the entropy.

It is convenient to work in terms of the functions
\be
E_L(x)=a(x)-2b(x)~~~~~~~~~~ E_H(x)=a(x)+2b(x)\;.
\ee
We now assume a compact spectrum, as appropriate to DSSYK model, and rewrite the integral equation for the density of staes in terms of its lower and upper bounds, 
\be
    \rho(E) = \int_0^1 dx\, \frac{\Theta(-(E-E_L(x))(E-E_H(x)))}{\pi\, \sqrt{-(E-E_L(x))(E-E_H(x))}}\;.
    \label{eq:DensOfStates}
\ee
First we assume monotonicity of the solution to the integral equation for the Lanczos coefficients (\ref{intdl}).\footnote{Recall that this integral equation applies to the ``bulk'' of the Lanczos spectrum, namely $a_n$ and $b_n$ for $n$ of $\mathcal{O}(L)$. For further comments on this assumption see \cite{Balasubramanian:2022dnj}. All models we have studied verified this assumption and DSSYK does as well, as shown numerically in Fig (\ref{numerical_LZ}).}  With this assumption, the integrand of (\ref{eq:DensOfStates}) only has support on $x$ for which $E$ lies within the range $[E_L(x_0),E_H(x_0)]=[a(x_0)-2b(x_0), a(x_0)+2b(x_0)]$.  So if some $E$ is outside the range $[E_L(x_0),E_H(x_0)]=[a(x_0)-2b(x_0), a(x_0)+2b(x_0)]$ for some $x_0$, then only $x<x_0$ can contribute  at that particular $E$. This implies that for $n\ll L$, still potentially much larger than $1$, the Lanczos coefficients $a_n$ and $b_n$ only depend on the density of states at the edges of the spectrum.

This means that for $E$ close to $E_L(0)$ (i.e., $E-E_L(0)\ll E_H(0)-E_L(0)$), $E$ must be close to $E_L(x)$ for the (small) range of $x$ that contributes to the integral. Continuity in $x$ gives us that $E_H(x)-E_L(x)$ is roughly equal to $E_H(0)-E_L(0)$ for small $x$, so the above integral can be approximated by
\be
    \rho(E)\approx \int_0^1 dx\, \frac{\Theta(E-E_L(x))}{\pi\, \sqrt{-(E-E_L(x))(E_L(x)-E_H(x))}} \approx \frac{1}{\pi \sqrt{E_H(0)-E_L(0)}} \int_0^1 dx\, \frac{\Theta(E-E_L(x))}{\sqrt{E-E_L(x)}}\;,
    \label{eq:approxdensity}
\ee
i.e., $E_H$ becomes only relevant as a scaling factor for finding $E_L$.   In other words (\ref{eq:approxdensity}) approximates (\ref{eq:DensOfStates}) for small $E$.  We now define
\be\label{udef}
u(E_L(x))\equiv\frac{1}{E_L'(x)}\;.
\ee
Using this function we can rewrite the density of states as a convolution \be\label{udefro}
    \rho(E) =\frac{1}{\pi \sqrt{E_H-E_L}} \int_{E_L(0)}^{E} dE_L\, \frac{u(E_L)\Theta(E-E_L)}{\sqrt{E-E_L}}\;.
\ee
Conversely, we may use this $u(E_L)$ to obtain 
\be\label{eq:xofL}
x(E_L)=\int_{E_L(0)}^{E_L}dE_L u(E_L) \,.
\ee
Given the definition of $E_L$ above, we can finally invert (\ref{eq:xofL})  to solve for $a(x)-2b(x)=E_L(x)$.

We now use a formula from  \cite{Balasubramanian:2023kwd}. There it was found that for the time evolution of the TFD, the stationary distribution in the Krylov basis at long times is given by
\be\label{pstfd2} 
\omega(x)= \frac{L}{Z_\beta}\, \int dE \,\frac{e^{-\beta E}}{\pi\,\sqrt{4b(x)^2-(E-a(x))^2}}\;.
\ee
Using again the functions $E_L(x),E_R(x)$, this can be rewritten as
\be 
\omega(x)=\frac{L}{Z_\beta} \int_{E_L(x)}^{E_H(x)} dE \,\frac{e^{-\beta E}}{\pi\,\sqrt{-(E-E_L(x))(E-E_H(x))}}\;.
\ee
At low temperatures, the distribution is concentrated at small $x$, where
\be
\omega(x)\approx\frac{L}{Z_\beta}\frac{1}{\pi \sqrt{E_H(0)-E_L(0)}}\int_{E_L(x)}^{\infty} dE \,\frac{e^{-\beta E}}{\sqrt{E-E_L(x)}}=\frac{Le^{-\beta E_L(x)}}{ Z_\beta \sqrt{\beta \pi(E_H(0)-E_L(0))}}\;.
\ee
We can then find the spread complexity plateau in terms of the function $u(E_L)$ defined above
\be\label{plateau_low_energy}
    \overline{x}=\int dx \frac{Lx e^{-\beta E_L(x)}}{ Z_\beta \sqrt{\beta \pi(E_H(0)-E_L(0))}}=\frac{L}{Z_\beta \sqrt{\beta \pi(E_H(0)-E_L(0))}}\int dE_L x(E_L) u(E_L) e^{-\beta E_L}\;.
\ee

\begin{figure}
    \includegraphics[width=1\linewidth]{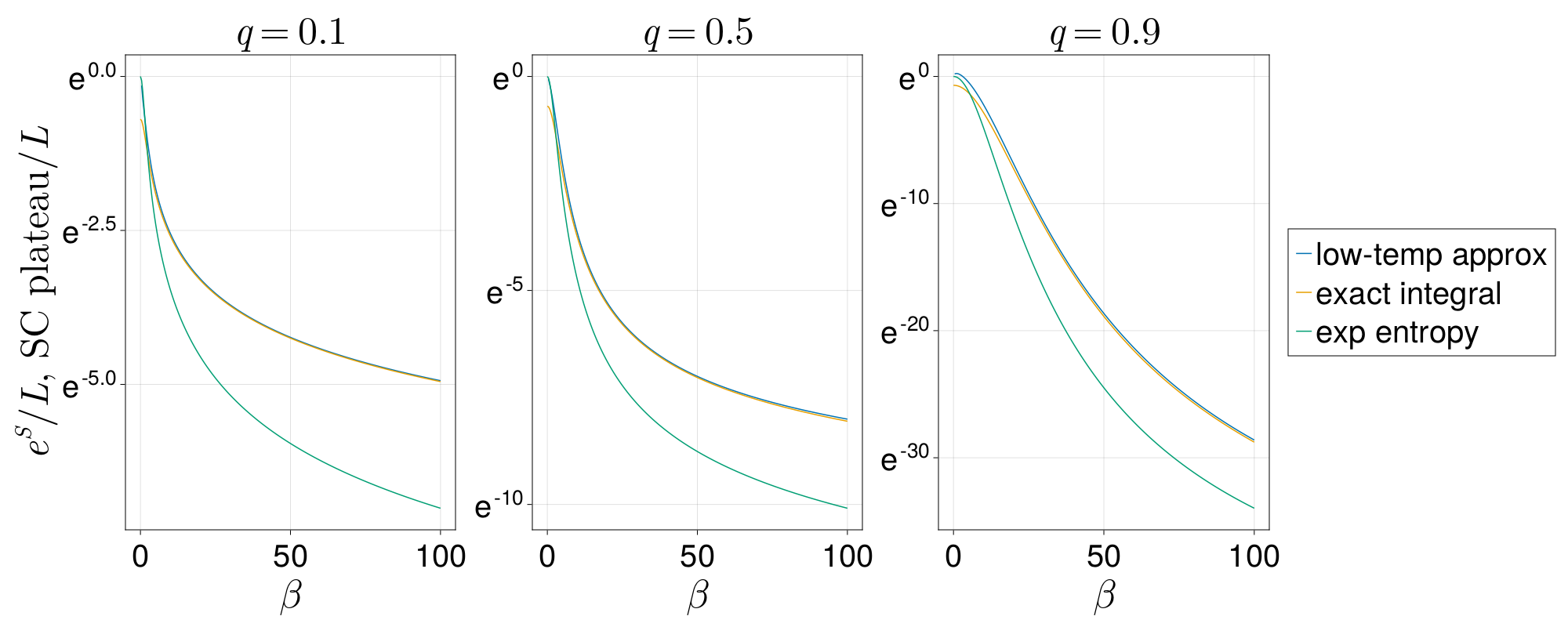}
    \caption{\label{lowerb} The plateau value of spread complexity associated to the TFD as a function of the temperature. Two nearly overlapping lines are shown: the exact result that takes into account the full density of states via \eqref{eq:DensOfStates} and \eqref{pstfd2} (orange), and the approximation that only uses the edge of the density of states \eqref{plateau_low_energy} (blue). For low temperatures these two match. We also plot the exponential of the entropy at the average thermal energy (green).     }
\end{figure}

This formula shows that the plateau value of spread complexity depends strongly only on the lower end of the density of states at low temperatures, and not on the details of density of states at higher energies aside from the cutoff $E_H(0)$. 
In Fig.~\ref{lowerb} we plot our analytical approximation to the plateau value for low temperatures, together with the exact large $L$ value given by formula (\ref{pstfd2}) and the exponential of the entropy of the SYK model at the appropriate temperature. The first two match perfectly as we lower the temperature, showing that the plateau value mostly depends on the lower end of the density of states at low temperatures. The exponential of the entropy, more precisely the density of states at the thermal energy, departs noticeably in this plot from the complexity plateau. In the next section we will study this deviation.

\section{Developments and applications}\label{IV}

In this section we will put our results in a broader context, focusing on three recent discussions in the literature: (a) the extension to higher dimensions, (b) the dynamical appearance of firewalls at exponentially late times in quantum gravity \cite{Stanford:2022fdt}, and (c) the Page curve for black hole information and the origin of black hole entropy.

\subsection{Extension to higher dimensions} 

We would like to extend our results to higher dimensional examples of the AdS/CFT correspondence. One key challenge is that CFTs have infinite dimensional Hilbert spaces, and time dependent quantities such as the spectral form factor and the TFD spread complexity, depend on the full spectrum. Equivalently, the Gibbs tails extend to arbitrary energies. 
To mitigate this, we can can focus on an energy window by convolving any  quantity with a  filter function $f(E/E_H)$ satisfying $f(0)=1$ and $f(x)=0$ for $x\geq 1$ (see \cite{Saad:2018bqo,Chen:2023hra} for analogous approaches). We can then consider initial states of the form
\be \label{Therf}
\vert\psi_{\beta}\rangle \equiv\frac{1}{\sqrt{Z_{\beta}}}\sum_n f(E/E_H) e^{-\frac{\beta E_n}{2}}\vert n,n\rangle\;,
\ee
where $\beta E_H\gg S(E_\beta)$, with $E_\beta$ the thermal energy defined by $S'(E_{\beta})=\beta$. Indeed a realistic black hole, e.g., formed by collapse, will have a wave function of the form (\ref{Therf}), with $f(E/E_H)\approx 1$ for energies of the same order as the thermal energy $E_\beta$, over an energy window around it, but then decaying rapidly to zero beyond that window. This ensures the initial state has a finite support in the energy basis.

We can analyze such more general situations using the tools we have already developed. Consider a density of states of the general CFT form
\be
\rho(E)=Ae^{c^k E^k}\;,
\ee
where $k\equiv\frac{d-1}{d}$ and $c$ has units of length. The dimensionful constant $c$ is related to the effective of number of degrees of freedom per thermal cell $c_{eff}$ as
$c\sim (c_{eff}V)^{\frac{1}{d-1}}$, where $V$ is the volume of the CFT. Below we assume  $c\gg \beta$.  If the system has large volume this is a standard thermodynamic limit. If the system is smaller than the thermal cell, then this limit forces $c_{eff}$ to be large. 

Instead of considering a particular filter function $f(E)$, we will just assume certain coarse grained properties associated with this function. Any particular function satisfying $f(E)= 1$ inside a window $E\in [E_L,E_H]$ will give rise to the same complexity plateau. This was the lesson of the previous section.  These coarse grained properties are $\beta (E_H-E_L)\gg 1$, where $E_H$ and $E_L$ are higher and lower cutoffs, beyond which the weighting function vanishes. We also assume $E_L(0)=0$ to simplify notation, although this is not very important either, as long as it is sufficiently small in comparison with the average energy.

We now first notice that for $cE_L \gg 1$, the function $u(E_L)$ defined in the previous section, see (\ref{udef}) and (\ref{udefro}), is
\be
u(E_L)\approx Ae^{c^k E_L^k}\sqrt{\pi E_H kc^k E_L^{k-1}}\;,
\ee
to leading order. Indeed, when we apply \eqref{udefro}, we have that the following integral is dominated by small $E_\delta \sim (kc^k E^{k-1})^{-1}$ as
\bea
    \rho(E)&=&\frac{1}{\pi \sqrt{E_H}} \int_{0}^{E} dE_\delta\, \frac{1}{\sqrt{E_\delta}}u(E-E_\delta)=\nonumber\\ 
    &\approx & \frac{1}{\pi \sqrt{E_H}} \int_{0}^{E} dE_\delta\, \frac{1}{\sqrt{E_\delta}} A e^{c^k E^k}e^{-kc^k E^{k-1}E_\delta}\sqrt{\pi E_H kc^k (E-E_\delta)^{k-1}} \nonumber\\
    &\approx & A\frac{e^{c^k E^k}}{\pi \sqrt{E_H}} \sqrt{\frac{\pi}{kc^k E^{k-1}}}\sqrt{\pi E_H kc^k E^{k-1}}
    = Ae^{c^k E^k} \;.
\eea
For the associated function defined below eq (\ref{udefro}) this gives\footnote{We recall that  $a(x)-2b(x)$ for small $x$ can be obtained from this function $x(E)$, see below eq (\ref{udefro}) and the definition of $E_L(x)$.} 
\bea
x(E_L)&=&\int_0^{E_L} Ae^{c^k E^k}\sqrt{\pi E_H kc^k E^{k-1}} dE=  \int_0^{E_L} Ae^{c^k (E_L-E_\delta)^k}\sqrt{\pi E_H kc^k (E_L-E_\delta)^{k-1}} dE_\delta\nonumber\\
&\approx & \int_0^{E_L} Ae^{c^k E_L^k}e^{-kc^k E_L^{k-1}E_\delta}\sqrt{\pi E_H kc^k (E_L-E_\delta)^{k-1}} dE_\delta=\frac{Ae^{c^k E_L^k}\sqrt{\pi E_H}}{\sqrt{kc^k E_L^{k-1}}}\;.
\eea 
The spread complexity plateau is then given by \eqref{plateau_low_energy}, evaluating to
\be
\overline{x}\approx \frac{L A^2\pi E_H}{Z_\beta \sqrt{\beta \pi E_H}}\int dE e^{2c^k E^k} e^{-\beta E}
\approx \frac{L A^2\pi E_H}{Z_\beta \sqrt{\beta \pi E_H}}\int dE e^{-\beta E+2c^k E^k}\;.
\ee
We can evaluate the last integral by steepest descent. There is a saddle point at 
\be
2kc^k E^{k-1}=\beta\implies E=(\frac{2kc^k}{\beta})^{1/(1-k)}\;.
\ee
This is consistent with $cE\gg1$. The second derivative at the saddle is $2k(k-1)c^k E^{k-2}=(k-1)(\frac{\beta^{2-k}}{2kc^k})^{1/(1-k)}$. We arrive at the following expression
\be
\overline{x}\approx \frac{L A^2\pi E_H}{Z_\beta \sqrt{\beta \pi E_H}} \sqrt{\frac{2\pi}{(1-k)(\frac{\beta^{2-k}}{2kc^k})^{1/(1-k)}}}e^{(2-2k)(2k\frac{c}{\beta})^{k/(1-k)}}\;.
\ee
The partition function $Z_\beta = AL\int dE e^{-\beta E+c^k E^k}$ can be treated anagolously to get
\be
Z_\beta\approx AL\sqrt{\frac{2\pi}{(1-k)(\frac{\beta^{2-k}}{kc^k})^{1/(1-k)}}}e^{(1-k) (k\frac{c}{\beta})^{k/(1-k)}}\;.
\ee
Most of this cancels out, leaving us with
\bea
\overline{x}\approx A\sqrt{\frac{\pi E_H}{\beta}}2^{1/(2(1-k))}
e^{(2^{1/(1-k)}-1)(1-k) (k\frac{c}{\beta})^{k/(1-k)}}\;.
\eea
In terms of the dimension $d$ of the CFT, namely $k=\frac{d-1}{d}$, this writes
\bea
\overline{x}\approx A\sqrt{\frac{\pi E_H}{\beta}}2^{d/2}
e^{\frac{2^{d}-1}{d} (\frac{d-1}{d}\frac{c}{\beta})^{d-1}}\;,
\eea
while in terms of the average energy $2kc^k E^{k-1}=\beta$ we have
\bea
\overline{x}=A\sqrt{\frac{\pi E_H}{\beta}}2^{d/2}
e^{\frac{2^{d}-1}{d}(\frac{1}{2})^{d-1} c^{k}\overline{E}^{k}}\;.
\eea
We conclude that the spread complexity plateau does not depend on $L$, the dimension of the Hilbert space.  Rather it only depends on the density of states at the energies of interest, namely the thermal energy in this case. We do see a mild dependence on the upper cutoff $E_H$ on the prefactor, but this is subleading with respect to the exponential. Notice that this formula implies that the spread complexity scales with $e^{O(S)}$, and not exactly $e^S$.

These results concern the value at saturation. We expect that the approach of spread complexity to saturation will display the  same features as for the DSSYK at long times (see Fig.\ref{AZ_SC}), namely dependence on the universality class of the time evolution.

\subsection{Black holes, white holes and firewalls}

A major open problem concerning the dynamics of quantum black holes is to understand whether an infalling observer sees a violation of the equivalence principle at the horizon. Such a violation can be broadly termed a ``firewall''. Arguments for the existence of firewalls were given in \cite{Mathur:2009hf,Almheiri:2012rt,Almheiri:2013hfa,Marolf:2013dba}. Arguments to save the equivalence principle were given in \cite{Bousso_2013,Verlinde_2013,Papadodimas:2012aq,Maldacena:2013xja,de_Boer_2019,Concepcion:2024nwv}.

Recently, \cite{Stanford:2022fdt} suggested an interesting dynamical argument for the appearance of firewalls at exponentially long times. Roughly, wormholes can describe tunneling processes between old black holes and white holes. The wormholes mediating the black hole to white hole transition are akin to those producing the ramp in the spectral form factor and thermal correlation functions \cite{Saad:2019lba,Saad:2018bqo,Saad:2019pqd}. 

There are some reasons to doubt this proposal. The first is that it seems to be basis dependent. For example, consider the basis of  microstates constructed in \cite{Balasubramanian:2022gmo,Balasubramanian:2022lnw,Climent:2024trz}.  These microstates, constructed via the Euclidean path integral, provide a basis for the states of the eternal black hole at $t=0$.  They can be evolved forward in time to give a basis in the black hole regime, and backward in time to give a basis in the white hole regime.  These states do not suffer from the classical problems suggesting the late appearance of firewalls in \cite{Stanford:2022fdt}. Of course, a superposition of these states may reveal a firewall, but it is not clear why this would happen. The second issue is that the probability expected for the black hole to white hole transition in \cite{Stanford:2022fdt} is $1/2$, and this does not seem to make a clear cut case that a firewall must the appear. For example, a spin up state in the $x$ basis is a linear superposition of up and down in the $z$ basis, with equal probability $p=1/2$ for each. But the $x$ basis state does not spontaneously transition unless it is perturbed in the appropriate basis. 

In JT gravity, this issue was explored in \cite{Iliesiu:2024cnh} by looking at the evolution of the wormhole size. In terms of this length, a transition from a black hole to  white hole regime should be seen as a transition from a growing ER bridge to a shrinking one.  This transition recalls the downward  slope from the complexity peak to the saturation value that appears in chaotic theories \cite{SpreadC}.  Indeed, Fig.~(\ref{AZ_SC}) we showed how this slope appears in the large $N$ SYK model. If we take spread complexity as a definition of the ER bridge size at a non-perturbative level, since the two quantities  already coincide in the perturbative regime, then the transition from the black hole regime to the white hole one becomes the transition to the downward complexity slope. This happens only for quantum chaotic theories since Poisson distributed spectra do not show the peak and slope \cite{SpreadC,Balasubramanian:2023kwd}.

\begin{figure}
    \includegraphics[width=0.8\linewidth]{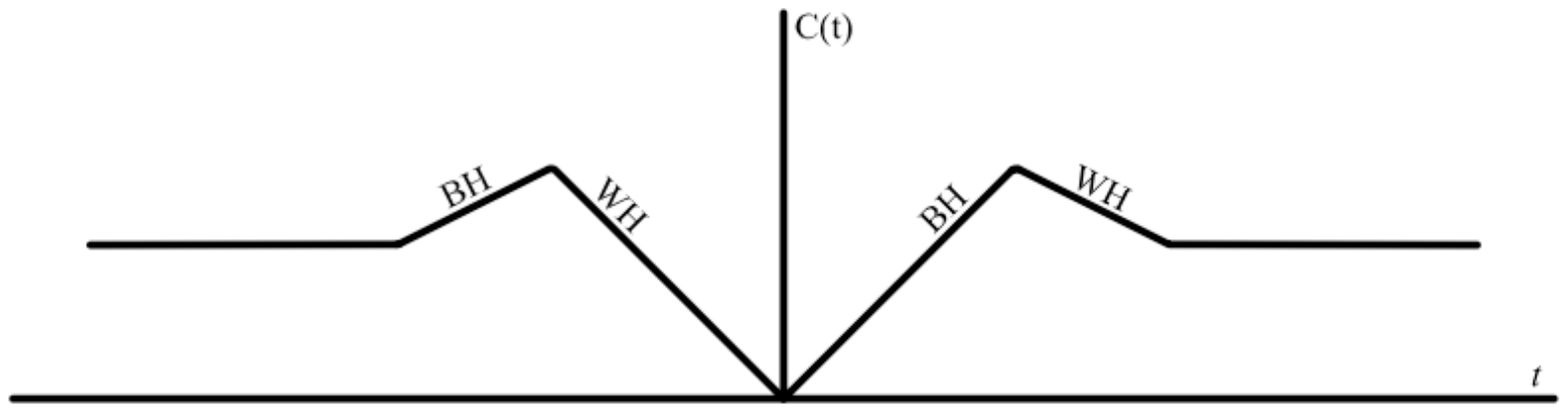}
    \centering
    \caption{\label{SpreadWH} The qualitative behavior of spread complexity for the thermofield double initial state evolving towards positive and negative times, assuming, e.g., a GUE universality class. Moving towards the future, the upward ramp codifies the black hole interior growth. At large times we transition to the downward complexity slope, which in the gravitational description should correspond to a white hole regime where the interior shrinks. At still larger times the spread complexity, and therefore the volume of ER bridge, saturates. The opposite behavior occurs at negative times with a period of interior growth before the classical white hole regime.}
\end{figure}

The universal behavior of spread complexity in, e.g., the GUE universality class, is depicted in Fig.~\ref{SpreadWH} for the thermofield double state. If we start at time equal to zero and move to the future, the ER volume grows linearly as expected. This is the classical black hole regime. But before saturating it decreases linearly for some time. This is the white hole regime appearing at exponentially long times. It can be dubbed a quantum white hole regime since it appears  microscopically because the Hilbert space has finite dimension, and also since it is requires the universal correlations of random matrix theory. Symmetrically, we can also evolve the intial state toward the past. Going to the past the spread complexity behaves in the very same way. It starts growing and stays that way for a very long time. Looking at this behavior from the far past to the future, this is the classical white hole regime. But further to the past we again get a complexity slope, this time interpreted as a quantum black hole regime, before reaching saturation. All of these regimes are depicted in Fig.~\ref{SpreadWH}.

Note that this dynamical structure happens here with probability one, i.e., this is the expected behavior of spread complexity. The origin of the complexity slope was explained in \cite{SpreadC} and further elaborated in \cite{Erdmenger:2023shk} -- it is rooted in the universal correlations of quantum chaotic systems. These correlations lead to a coherent evolution  of the TFD wavefunction on the Krylov chain, so that it overshoots the plateau value, and then has to decrease after bouncing back from a peak.  But notice that although there is a white hole regime where the interior shrinks, this does not make the interior sufficiently young to suffer from the problems mentioned in  \cite{Stanford:2022fdt}. The black hole expends most of its time being large.

\subsection{Black hole entropy and the Hilbert space of quantum gravity}

The problem of constructing finite dimensional Hilbert spaces for black holes in quantum gravity has seen significant development in recent years \cite{Penington:2019kki,Hsin:2020mfa,Chandra:2022fwi,Balasubramanian:2022gmo,Balasubramanian:2022lnw,Boruch:2023trc,Antonini:2023hdh,Climent:2024trz,Iliesiu:2024cnh,Boruch:2024kvv}. These Hilbert spaces have dimension consistent with the Bekenstein-Hawking entropy, and are constructed in terms of basis elements that are, in some sense, generalized coherent states since they have semiclassical descriptions and  are not fully orthogonal with respect to each other.

Here, we have provided a novel way of approaching this problem.  To understand this, consider the family of black hole microstates proposed in \cite{Papadodimas:2015xma} and later explained in the present context in \cite{Banerjee:2023liw}. This is the family of time evolved thermofield double states, namely
\be 
\vert \psi_{\beta+2it}\rangle = e^{-iHt}\vert \psi_{\beta}\rangle\;.
\ee
This is a continuous set of microstates, naively too many to constitute a finite dimensional black hole basis, especially in a holographic field theory where they correspond to different slices of the same eternal black hole. 
However, we can in fact construct a basis from these states.\footnote{This has been noticed as well recently from different perspectives in Refs. \cite{Magan:2024nkr,Banerjee:2024fmh}.} To show this  first notice that
\be
\langle \psi_{\beta} (t')\vert \psi_{\beta}(t)\rangle =\frac{Z_{\beta+i(t-t')}}{Z_\beta}\;,
\ee
where $Z_\beta$ is the partition function. If the phases are incoherent relative to each other as we expect in a chaotic theory, the overlap will cancel as
$\Delta t\equiv t-t'\rightarrow\infty$ because we will be summing up random phases.
This is easily verified by taking a continuous approximation for the density of states $\rho(E)$ and the integral representation of the partition function. Similarly, this can also been seen in RMT when we use the average density of states, e.g., for a Gaussian random matrix theory the density of states is the semicircle law and the analytically continued partition function decays as a Bessel function.

The fact that $\langle \psi_{\beta} (t')\vert \psi_{\beta}(t)\rangle$ goes to zero as $\Delta t\equiv t-t'\rightarrow\infty$ suggests that by taking a discrete set of $\Omega$ states of the form
\be \label{basisom}
\vert\psi_{\beta} (t_1)\rangle\,\,\,\,\,\,\,\, \vert\psi_{\beta}(t_2)\rangle\,\,\,\,\,\,\,\,\vert\psi_{\beta}(t_3)\rangle\,\,\,\,\,\,\,\,\cdots\,\,\,\,\,\,\,\,\vert\psi_{\beta}(t_{\Omega})\rangle\;,
\ee
we should get a basis for a Hilbert space of dimension $\Omega$. Of course within  any finite dimensional microcanonical window if $\Omega (E)>e^{S(E)}$ then the basis will be overcomplete. To ascertain that the dimension of the Hilbert space gets appropriately corrected at this threshold, we can compute the rank of the matrix of overlaps between these states. This matrix is defined here as
\be 
G_{i j}\equiv \langle \psi_{\beta} (t_i)\vert \psi_{\beta} (t_j)\rangle\;,
\ee
where $i,j=1,\cdots, \Omega$. Although we cannot compute this matrix exactly unless we have all the information about the spectrum of the theory, we can compute  statistical aspects of $G$ for chaotic theories. Let is compute the $n$th moment
\be 
G_{1 2}G_{2 3}\cdots G_{\Omega 1}=\langle \psi_{\beta} (t_1)\vert \psi_{\beta} (t_2)\rangle\langle \psi_{\beta} (t_2)\vert \psi_{\beta} (t_3)\rangle\cdots \langle \psi_{\beta} (t_n)\vert \psi_{\beta} (t_1)\rangle\;.
\ee
Assuming $t_1$ is sufficiently large and that $t_j=j t_1$, with $j=1,\cdots , \Omega$, all the time differences $t_i-t_j$ are large. Therefore this quantity can be estimated by computing the long time average of each of the time variables. One obtains
\be 
G_{1 2}\cdots G_{\Omega 1}=\lim_{T_1 ,\cdots,  T_n\to\infty} \left(\prod_{k=1}^{n}\frac{1}{T_k}\int_{-T_k/2}^{T_k/2} dt_k \right) \frac{1}{Z_{\beta}^n}\sum_{j_1,\cdots j_n}\,e^{-\beta\sum_{l=1}^{n}E_{j_l}-i t_l (E_{j_{l+1}}-E_{j_l})}=\frac{Z(n\beta)}{Z_{\beta}^n}\;.
\ee
This precisely reproduces the universal statistics found in \cite{Balasubramanian:2022gmo,Balasubramanian:2022lnw,Climent:2024trz} for black hole microstates constructed with heavy shells behind horizon. As shown in those papers, an overlap matrix with this property implies that the dimension of the  microcanonical Hilbert space expanded by ~(\ref{basisom}) is equal to $\Omega$ if $\Omega$ is smaller than the microcanonical degeneracy $e^{S(E)}$, and is equal to the microcanonical degeneracy $e^{S(E)}$ if $\Omega$ is larger than $e^{S(E)}$.

Notice that for $n=2$ the quantity we just computed is equal to the plateau of the spectral form factor. We see here that the problem of deriving the saturation plateau of the spectral form factor is basically the same problem as that of proving that the TFD time evolved states are a basis of the Hilbert space with the right dimensionality.

The above discussion relates recent constructions of Hilbert spaces for black holes in quantum gravity with the older considerations based on time evolved thermofield double states. The relation with the approach in the present paper is that the Krylov basis associated with an initial thermofield double state is one in which, by construction, we can expand the time evolved thermofield double. So there is a unitary transformation that performs a change of basis from (\ref{basisom}) to the Krylov basis. This unitary might be difficult to construct explicitly, but the fact that the time-evolved thermofield double basis is finite implies the Krylov basis is finite and vice-versa. For the Krylov basis, as we showed above, the finiteness of the basis is controlled by the vanishing of the off-diagonal Lanczos spectrum, and this vanishing is ensured by the integral equation~(\ref{intdl}) or by the saddle point equations (\ref{sadlan}).

\section{Discussion}\label{V}

The construction of finite dimensional Hilbert spaces for black holes in quantum gravity is an important direction in the field. In this article we have provided a new approach for such constructions via  Krylov subspace methods. We started by applying this construction to the Double Scaled SYK model, where we showed how to construct a finite tridiagonal Hamiltonian with the right moments. This improves on previous approaches where this matrix was constructed in an approximation where it was taken to be infinite dimensional. The latter approximation effectively extrapolates a sub-exponential part of the full Hilbert within which the TFD state has a good classical gravity description.  By contrast, we reconstruct the entire Hilbert space, including the parts that are explored at exponentially late times, and where quantum effects lead to saturation of the size of the Einstein-Rosen bridge in the gravitational description.  Our approach does not require any doubled scaled limit, and  can be applied to more realistic theories of quantum gravity, such as holographic CFT's in higher dimensions. The key input is the density of states of the theory in the thermodynamic limit, which follows directly from the Bekenstein-Hawking entropy. As a byproduct, in these finite dimensional Hilbert spaces, the ER bridges are bound to saturate at exponentially late times, and we can access interesting dynamics in the approach to saturation. These dynamics include a transition to a white hole regime, at least in systems that manifest underlying quantum chaos.  This finding  might have implications for the fate of observers falling into a black hole.

In the future, it would be useful to show precisely whether the semiclassical limit of spread complexity is holographically dual to the Einstein-Rosen bridge volume for the AdS/CFT correspondence in higher dimensions. Likewise it would be interesting to understand whether these ideas can be applied in the matrix model of M-theory \cite{Banks:1996vh,Balasubramanian:1997kd,Polchinski:1999br} or in the chiral theories dual to extremal black holes \cite{Balasubramanian:2003kq}. It would also be interesting to better understand the implications of the Lanczos descent on  gravitational dynamics, in particular in the black hole interior. As part of this, it is important to ascertain what basis is best suited for the analysis of infalling observers. Confusingly, as discussed in the previous section, it seems that different basis states suggest different interpretations (e.g., a firewall vs. smooth horizon crossing).   Perhaps there is a physical reason for a choice of one these options, or perhaps these are complementary but equivalent descriptions of the same phenomena. The Krylov basis is certainly canonical, and optimal in several respects, but it is far from clear that this basis is the most illuminating for analysis of the infalling observer.

\paragraph{Acknowledgements: }  We are grateful to Pawel Caputa for many discussions about spread complexity and the Lanczos approach, and Martin Sasieta for conversations about quantum chaos and the black hole interior. We also wish to thank the participants of the workshop ``The microscopic origin of black hole entropy'', held in the Aspen Centre for Physics (ACP), for discussions on DSSYK and low dimensional gravity. J.M wishes to thank Guido Van der Velde and Mario Solis for initial discussions on the Lanczos spectrum of DSSYK. This work was performed in part at the ACP, which is supported by a grant from the Simons Foundation (1161654, Troyer). The work of VB and QW is supported by a DOE through DE-SC0013528 and QuantISED grant DE-SC0020360. VB is supported in part by the Eastman Professorship at Balliol College, Oxford. 
 The work of JM is supported by CONICET, Argentina. J.M acknowledges hospitality and support from the International Institute of Physics, Natal, through Simons Foundation award number 1023171-RC. PN is supported by a Fulbright-Nehru Postdoctoral Research Fellowship. She
would like to thank hospitality of the University of Pennsylvania
during the course of this work.

\newpage

\appendix
\section{Symmetries of ensembles of random matrix theories}\label{appendixA}

The symmetries in Table 1 can be derived by finding antihermitian matrices $M$ such that $[M,H]$ has the same block structure as $H$.

For example, lets begin with a block structure $H=\begin{bmatrix}0&Z\\Z^\dagger &0\end{bmatrix}$. The commutator must give us the same block structure as the original matrix
\be
\left[\begin{bmatrix}A&B\\C&D\end{bmatrix}, \begin{bmatrix}0&Z\\Z^\dagger &0\end{bmatrix}\right] = \begin{bmatrix}BZ^\dagger-ZC & AZ-ZD\\DZ^\dagger-Z^\dagger A &CZ-Z^\dagger B\end{bmatrix}=\begin{bmatrix}0&Z'\\Z'^\dagger &0\end{bmatrix}.
\ee
This can only happen for all $Z$ if $B=C=0$, $A=-A^\dagger$, $D=-D^\dagger$. Exponentials of antihermitian matrices give unitaries, giving us symmetries of the form $\begin{bmatrix}
    U_1&0\\0&U_2
\end{bmatrix}$. 

Similar results also hold for the equivalent for real and quaternion matrices; $\begin{bmatrix}0&A\\A^T &0\end{bmatrix}$ retains its block structure when conjugated with orthogonal blocks $\begin{bmatrix}
    O_1&0\\0&O_2
\end{bmatrix}$, and $\begin{bmatrix}0&B\\B^\dagger &0\end{bmatrix}$ with quaternion unitary blocks $\begin{bmatrix}
    S_1&0\\0&S_2
\end{bmatrix}$.

Next, for the block structure $\begin{bmatrix}A&B\\\overline{B} &-\overline{A}\end{bmatrix}$, we want
\be
\left[\begin{bmatrix}E&F\\G&H\end{bmatrix}, \begin{bmatrix}A&B\\\overline{B} &-\overline{A}\end{bmatrix}\right] = \begin{bmatrix}EA+F\overline{B}-AE-BG & EB-F\overline{A}-AF-BH\\GA+H\overline{B}-\overline{B}E+\overline{A}G &GB-H\overline{A}-\overline{B}F+\overline{A}H\end{bmatrix}=\begin{bmatrix}A'&B'\\\overline{B}' &-\overline{A}'\end{bmatrix}.
\ee
Setting the two $A'$ blocks equal gives $G=-\overline{F}=Y$ and $E=\overline{H}=X$, which is enough to make the $B'$ blocks equal as well, and the symmetry would be any exponential $\exp\left(\begin{bmatrix}X&-\overline{Y}\\Y&\overline{X}\end{bmatrix}\right)$.

For the single block complex $M$ satisfying $M^\dagger=M$ and $M^T=-M$, we need $[A,M]^\dagger=[A,M]$ (so $A^\dagger = -A$) and $[A,M]^T=-[A,M]\implies M^TA^T-A^TM^T=A^TM-MA^T=-(AM-MA)$, so $A^T=-A$, so $A$ is skew symmetric and real. Exponentiating this gives an orthogonal symmetry group.

Next for the $\begin{bmatrix}0&Z\\\overline{Z} &0\end{bmatrix}$ block structure, we have
\be
\left[\begin{bmatrix}A&B\\C&D\end{bmatrix}, \begin{bmatrix}0&Z\\\overline{Z} &0\end{bmatrix}\right] = \begin{bmatrix}B\overline{Z}-ZC & AZ-ZD\\D\overline{Z}-\overline{Z} A &CZ-\overline{Z} B\end{bmatrix}=\begin{bmatrix}0&Z'\\\overline{Z}' &0\end{bmatrix}.
\ee
This can only happen if $B=C=0$ and $A=\overline{D}$. The condition $Z^T=Z$ and $Z'^T=Z'$ implies that $ZA^T-D^TZ=AZ-ZD$, so $A=-D^T$ as well. This implies that both $A,D$ are anti-hermitian, so the exponential of $A$ is some unitary $U_1$ while the exponential of $D$ must be its conjugate $\overline{U}_1$, giving us symmetry by $\begin{bmatrix}
    U_1&0\\0&\overline{U}_1
\end{bmatrix}$

Lastly, for the block structure $\begin{bmatrix}0&Y\\-\overline{Y} &0\end{bmatrix}$, we have
\be
\left[\begin{bmatrix}A&B\\C&D\end{bmatrix}, \begin{bmatrix}0&Y\\-\overline{Y} &0\end{bmatrix}\right] = \begin{bmatrix}-B\overline{Y}-YC & AY-YD\\-D\overline{Y}+\overline{Y} A &CY+\overline{Y} B\end{bmatrix}=\begin{bmatrix}0&Y'\\-\overline{Y}' &0\end{bmatrix}.
\ee
This again can only happen if $B=C=0$ and $A=\overline{D}$. The condition $Y^T=-Y$ and $Y'^T=-Y'$ implies that $-YA^T+D^TY=-(AY-YD)$, which again implies $A=-D^T$ is anti-hermitian. So we have the same symmetry $\begin{bmatrix}
    U_1&0\\0&\overline{U}_1
\end{bmatrix}$.

\bibliographystyle{utphys}
\bibliography{main}

\end{document}